\documentclass[11pt]{article}

\usepackage{epsf}
\usepackage{graphicx}% Include figure files
\usepackage[titletoc,title]{appendix}

\newcommand{\lsim}{\raisebox{0.6mm}{$\, <$} \hspace{-3.0mm}\raisebox{-1.5mm}{\em $\sim \,$}}
\newcommand{\gsim}{\raisebox{0.6mm}{$\, >$} \hspace{-3.0mm}\raisebox{-1.5mm}{\em $\sim \,$}}

\newcommand{\beq}{\begin{equation}}   
\newcommand{\eeq}{\end{equation}}
\newcommand{\bea}{\begin{eqnarray}}   
\newcommand{\eea}{\end{eqnarray}}
\def\GEV#1{10^{#1}{\rm\,GeV}}

\def\gev{{\rm\,GeV}}

\def\mgluino{m_{\tilde{g}}}

\def\lrf#1#2{ \left(\frac{#1}{#2}\right)}
\def\lrfp#1#2#3{ \left(\frac{#1}{#2} \right)^{#3}}

\def\stau{\tilde{\tau}}
\def\cos{{\rm cos}}
\def\sin{{\rm sin}}
\def\tan{{\rm tan}}
\let\bar\overline 
\let\tilde\widetilde

\begin{document}
\baselineskip=18pt

\begin{titlepage}

\begin{flushright}
UT--11--16\\
IPMU--11--0088\\
\end{flushright}

\vskip 1.35cm
\begin{center}
{\Large \bf
LHC signature with long--lived stau in high reheating temperature scenario
}
\vskip 1.2cm
Motoi Endo$^{1,2}$, Koichi Hamaguchi$^{1,2}$, Kouhei Nakaji$^{1}$
\vskip 0.4cm

{\it $^1$  Department of Physics, University of Tokyo,
Tokyo 113-0033, Japan\\
$^2$ Institute for the Physics and Mathematics of the Universe, 
University of Tokyo,\\ Chiba 277-8568, Japan
}

\vskip 1.5cm

\abstract{We study the possibility of observing the stau signal at LHC
 in case that the gravitino is the lightest supersymmetric particle
  and the stau is the next lightest supersymmetric particle in
  high reheating temperature scenario. We show that a number of stau signals can be observed at LHC
 for $\sqrt{s}=7~{\rm TeV}$ and an integrated luminosity of $1~{\rm fb}^{-1}$
   in most of the parameter region for the reheating temperature $T_R\gsim{\cal O}(10^8)~{\rm GeV}$.
    We also show that the  parameter region with $T_R\gsim2\times10^9~{\rm GeV}$,    
    which is consistent with the thermal leptogenesis, is all covered for
   $\sqrt{s}=14~{\rm TeV}$ with $10~{\rm fb}^{-1}$.

}
\end{center}
\end{titlepage}

\setcounter{footnote}{0}
\setcounter{page}{2}

%%%%%%%%%%%%%%%%%%%%%%%%%%%%%%%%%%%%
\section{Introduction}
%%%%%%%%%%%%%%%%%%%%%%%%%%%%%%%%%%%%
Supersymmetry (SUSY) is one of the most plausible models beyond the standard model. 
In some classes of minimal supersymmetric standard model (MSSM) with the gravitino,
the gravitino ($\tilde{G}$) becomes the lightest supersymmetric particle (LSP) and the lightest
stau ($\stau$) becomes the next-to-lightest supersymmetric particle (NLSP). 
Such a scenario may predict the stau lifetime longer than
${\cal O}$(1) sec  because of the weak coupling between the gravitino and the stau. 
The existence of the long-lived charged massive particles like the stau is appealing
from the viewpoint of the discovery at the collider, 
since they are observed as charged tracks at the detector. 
The search for such long-lived particles at LHC has already started~\cite{LHCCHAMP}.  

However, there are some cosmological problems 
in this scenario.
The present energy density of the gravitino produced in the
thermal scattering at the epoch of reheating may exceed the observed dark matter energy density. Moreover, the existence or the late-time decay of the stau at the epoch of Big Bang Nucleosynthesis (BBN)
may spoil the successful BBN. It has been pointed out that,
from these two cosmological constraints,
upper bounds on the gluino mass are obtained for a given reheating temperature and for a given stau mass~\cite{FIY} (see also
\cite{RandS}).  Such upper bounds on the gluino mass predicts promising
collider signatures of this long-lived stau scenario.

In this paper, we study the observability of the stau signals at LHC in this long-lived stau scenario 
under the cosmological constraints, assuming that the R-parity is conserved, and that
there is no entropy production after the reheating.  
We show that a higher reheating temperature predicts a lower gluino mass,
and therefore more stau signals can be observed at LHC. 
It is shown that a number of stau signals can be observed 
at the early LHC (the center-of-mass energy $\sqrt{s}=7~{\rm TeV}$ and the integrated luminosity $L_i=1~{\rm fb}^{-1}$)
in most of the parameter region for $T_R\gsim{\cal O}(10^8)~{\rm GeV}$.  
    We also show that the  parameter region with $T_R\gsim2\times10^9~{\rm GeV}$,    
    which is consistent with the thermal leptogenesis~\cite{Fukugita:1986hr,lepto}, is all covered 
    for $\sqrt{s}=14~{\rm TeV}$ and $L_i=10~{\rm fb}^{-1}$.
This paper is the complete version of our previous work \cite{Endo:2010ya}.
In addition to the analysis  in our previous paper, we include the followings in this paper.
We study the parameter region where the stau annihilates 
near the pole of CP-even heavy Higgs boson. 
The upper bounds on the  gluino mass are shown in various gaugino mass relations. 
The results using the detector simulation are shown in more details.

This paper is organized as follows. In Section \ref{cosmology},
we discuss the cosmological constraints. 
In Section 3, the constraints from Tevatron and the signatures at LHC are studied. We conclude this paper in Section 4.

%%%%%%%%%%%%%%%%%%
\section{Cosmological Constraints}
%%%%%%%%%%%%%%%%%%
\label{cosmology}
In this section, we discuss the cosmological constraints in the scenario with a long lived stau
and with high reheating temperature. There are mainly two constraints: the constraints from
gravitino over-production and the constraints from BBN. The over-production bound is reviewed 
in Sec.\ref{sec:Omega32}, where an upper bound on the gluino mass is obtained
for a given gravitino mass and a reheating temperature.
In Sec.\ref{sec:BBN}, we discuss the BBN bound, and 
we get an upper bound on the gravitino mass for a given stau mass. Combining these two cosmological constraints, an upper bound on the gluino mass is obtained for a given stau mass and
a reheating temperature.

We discuss the three different annihilation processes of staus: (i) electroweak processes, (ii) annihilation into light Higgs bosons($h$) \cite{enhance1,enhance2} and (iii) annihilation near the pole of CP-even heavy Higgs boson($H$)
\cite{enhance2}. 

\subsection{Gravitino Over-production Bound}
\label{sec:Omega32}
%%%%%%%%%%%%%%%%%%%%%%%%%%%%%%%%%
The gravitinos are produced by the scattering process of particles in thermal bath after the epoch of reheating~\cite{thermaltr,Bolz:2000fu,Pradler:2006hh,Rychkov:2007uq}. The gravitino abundance takes the form of \cite{Pradler:2006hh}
\begin{eqnarray}\label{Ythermal0}
\Omega_{3/2}h^2 &\simeq&
\lrf{T_R}{\GEV{8}}
\left(3.7\times 10^{-4}\left(\frac{m_{3/2}}{100~{\rm GeV}}\right)\right.\nonumber\\
&&
\left. +
\lrf{1\gev}{m_{3/2}}\left[ 
0.14\lrfp{m_{\tilde{B}}}{1{\rm\,TeV}}{2}
+
0.38\lrfp{m_{\tilde{W}}}{1{\rm\,TeV}}{2}
+
0.34\lrfp{\mgluino}{1{\rm\,TeV}}{2}
\right]\right)
,
\label{gluinomass}
\end{eqnarray}
where $m_{3/2}$, $m_{\tilde{B}}$, $m_{\tilde{W}}$ and $m_{\tilde{g}}$ are the physical masses of the gravitino, 
the bino, the wino and the gluino respectively. 
The reheating temperature $T_{R}$ is defined by
\begin{eqnarray}\label{eq:T_R}
T_R &=& \lrfp{\pi^2 g_*(T_R)}{90}{-1/4}\sqrt{\Gamma_\phi M_P},
\end{eqnarray}
where $\Gamma_\phi$ is the inflaton decay rate,  $g_*$ is effective degrees of freedom and $M_P=2.4\times 10^{18}$ GeV is the Planck scale.
We used the one--loop renormalization group equations
to evolve the running masses of gauginos up to the scale $\mu = T_R$. 
The numerical coefficients, which depend on $T_{R}$ logarithmically, are evaluated at  $T_{R}=10^8 {\rm GeV}$ in Eq.(\ref{Ythermal0}). 
In the numerical analysis, we include those logarithmic dependences.
Note that Eq.(\ref{Ythermal0}) potentially includes an ${\cal O}(1)$ uncertainty
\cite{Bolz:2000fu,Rychkov:2007uq}.

Although gravitinos can be also produced by inflaton decay \cite{inflatondecay} and the moduli decay \cite{modulidecay}, 
they are model dependent, and we do not include these production processes, for simplicity. 
In the present scenario, late-time stau decay also produces gravitinos,
\begin{eqnarray}
\Omega^{NT}_{3/2}h^2\simeq2.8\times 10^{-7}\left(\frac{Y_{\stau}}{10^{-15}}\right)
\left(\frac{m_{3/2}}{1{\rm GeV}}\right),
\end{eqnarray}
where $Y_{\stau}\equiv n_{\stau}/s$ is the stau abundance, $n_{\stau}$=$n_{\stau}^{+}+n_{\stau}^{-}$ is stau number density after the freeze-out, and $s$ is the entropy density.
However, its contribution is negligible in the parameter region of our interest, and we neglect it in the following discussion.

\begin{figure}[t]
\begin{center}
\includegraphics[width=12cm]{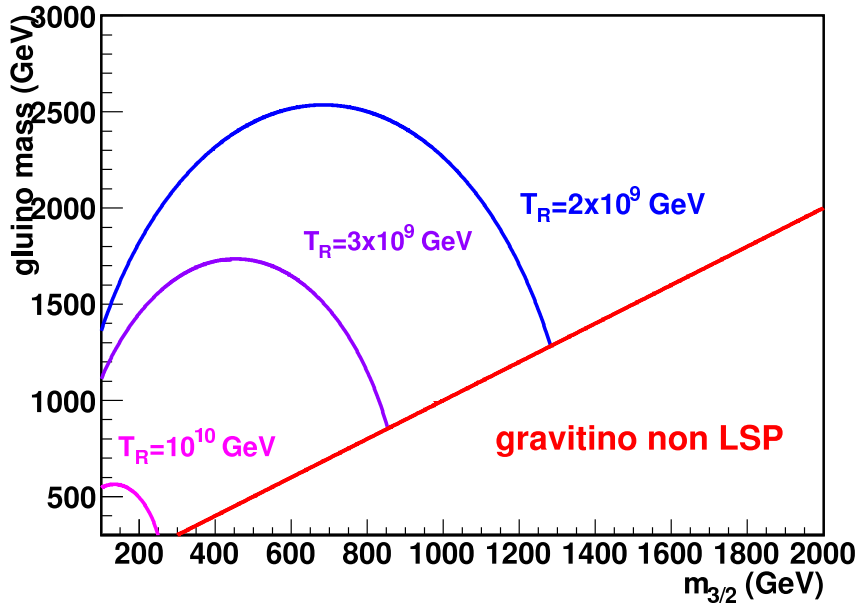}
\end{center}
\caption{The upper bound on the gluino mass in case that $M_{\tilde{B}}=
M_{\tilde{W}}=m_{3/2}.$}
\label{fig:gralsp}
\end{figure}

The gravitino energy density should not exceed the observed DM density \cite{darkmatterdensity},
\begin{eqnarray}
\Omega_{3/2}h^2\leq 0.122~(95\% {\rm C.L.}).
\label{Omegah}
\end{eqnarray}
This constraint gives the upper bound on the value of gluino mass for a given gravitino mass and
reheating temperature, for fixed values of bino mass and wino mass. 
The most conservative bound is obtained for $m_{\tilde{B}}=m_{\tilde{W}}=m_{3/2}$, 
which is shown in Fig.\ref{fig:gralsp}.
In the lower gravitino mass region, the term proportional to $m_{\tilde{g}}^2/m_{3/2}$  
dominantly contributes to the right-hand side in Eq.(\ref{Ythermal0}). Thus, in that
region, the upper bound on $m_{\tilde{g}}$ is proportional to $\sqrt{m_{3/2}}$,
as we can see in Fig. \ref{fig:gralsp}.
On the other hand, we see that the upper bounds drop down
in the higher gravitino mass region, because
the terms proportional to  $m_{3/2}$ and $m_{\tilde{B}/\tilde{W}}^2/m_{3/2}$ become
non-negligible.

We find that in gravitino LSP scenario,
 the gluino mass cannot be higher than $2.5~{\rm TeV}$ for 
$T_R\gsim 2\times 10^{9}~{\rm GeV}$ which is required in the 
thermal leptogenesis \cite{Fukugita:1986hr,lepto}. 
It is also found that the  reheating temperature higher than ${\cal O}(10^{10})~{\rm GeV}$
is not allowed in the scenario with gravitino LSP.
Note that these bounds are the most conservative ones and
should be satisfied independently of the BBN constraints, unless there is an entropy production after the gravitino production.
%%%%%%%%%%
%%%%%%%%%
%%%%%%%%%%

%%%%%%%%%%%%%%%%%%%%%%%%%%%%%%%
\subsection{BBN Bound}
\label{sec:BBN}
%%%%%%%%%%%%%%%%%%%%%%%%%%%%%%%
In the present scenario, staus are long-lived so that they may affect BBN.
There is an upper bound on the stau lifetime 
depending on the value of stau relic abundance at the BBN epoch.
The stau lifetime is given by
\begin{eqnarray}
\stau_{\tilde \tau} = \frac{48\pi M_{\mathrm P}^2 m_{3/2}^2}{m_{\tilde \tau}^5}
\left(1-\frac{m_{3/2}^2}{m_{\tilde \tau}^2}\right)^{-4},
\end{eqnarray}
and hence the upper bound on the stau lifetime can be translated into an
upper bound on the gravitino mass for a given stau mass.

Among the constraints on stau lifetime from BBN, the most stringent bound comes from 
the catalyzed effect where stau forms a bound state with $^4{\rm He}$ and 
overproduces $^6{\rm Li}$ \cite{cbbn1}.
The constraint is roughly $\stau_{\tilde \tau}<1000~$sec for $Y_{\stau}\gsim10^{-15}$, while 
there is almost no constraint on stau lifetime for $Y_{\stau}\lsim 10^{-15}$. 
It is also important for large $Y_{\stau}$ 
 that the energetic hadrons produced by stau-decay modifies the abundance of D nuclei
 \cite{bbn2}.

Stau abundance $Y_{\stau}$ depends on the value of the stau annihilation cross section, 
and hence on the stau annihilation process. 
In this paper, we consider the following three different parameter regions.
\begin{enumerate}
\item[(A)] In most of the parameter region of the MSSM, the stau annihilation is dominated by the electroweak interaction. 
\item[(B)] When the $\stau-\stau-h$ coupling is large, the stau annihilation into the light CP-even Higgs bosons $h$ is enhanced~\cite{enhance1,enhance2}.
\item[(C)] When the $\stau-\stau-H$ coupling is large and the heavy Higgs mass satisfies the condition $m_H\simeq 2m_{\stau}$, the staus can annihilate at the resonance of the heavy Higgs boson $H$~\cite{enhance2}. 
\end{enumerate}
In the cases of (B) and (C), the stau abundance $Y_{\stau}$ can be significantly reduced compared to the case (A)~\cite{enhance1,enhance2}.
In the following subsections \ref{sec:A}--\ref{sec:C}, we discuss the gravitino mass upper bound for a given stau mass in these three different parameter regions. The obtained BBN bound on the gravitino mass is combined with the over-production bound discussed in Sec.\ref{sec:Omega32}, which leads to the  gluino mass upper bound for a
given stau mass and reheating temperature.
 
%%%%%%%%%%%%%%%%%%%%%%%%%%
\subsubsection{(A) Stau Annihilation via Electroweak process}
\label{sec:A}
%%%%%%%%%%%%%%%%%%%%%%%%%%%%
When the electroweak process is dominant, the stau abundance is estimated as 
\cite{Asaka:2000zh}
\begin{eqnarray}
  Y_{\stau} \simeq 7 \times 10^{-14} 
  \times
  \left( \frac{m_{\stau}}{100 \rm{GeV}} \right).
  \label{eq:stau-electroweak}
\end{eqnarray}
Then, the bound from BBN on gravitino mass is \cite{bbn2}
\begin{eqnarray}
  m_{3/2} \lsim 
  \left\{ \begin{array}{ll}
  0.4 \gev - 10 \gev & (100{\rm GeV}<m_{\stau}<450{\rm GeV}) \\
  10 \gev - 20 \gev & (450{\rm GeV}<m_{\stau}<1000{\rm GeV})
\end{array} \right.,
\end{eqnarray}
where the constraint in the region $100{\rm GeV}<m_{\stau}<450{\rm GeV}$ comes from the bound
on $^6{\rm Li}$ overproduction by the catalyzed effect, whereas the bound on the hadronic decay of staus gives the constraint in the region $450{\rm GeV}<m_{\stau}<1000{\rm GeV}$.

%%%%%%%%%%
\begin{figure}[t]
\begin{center}
\includegraphics[width=12cm]{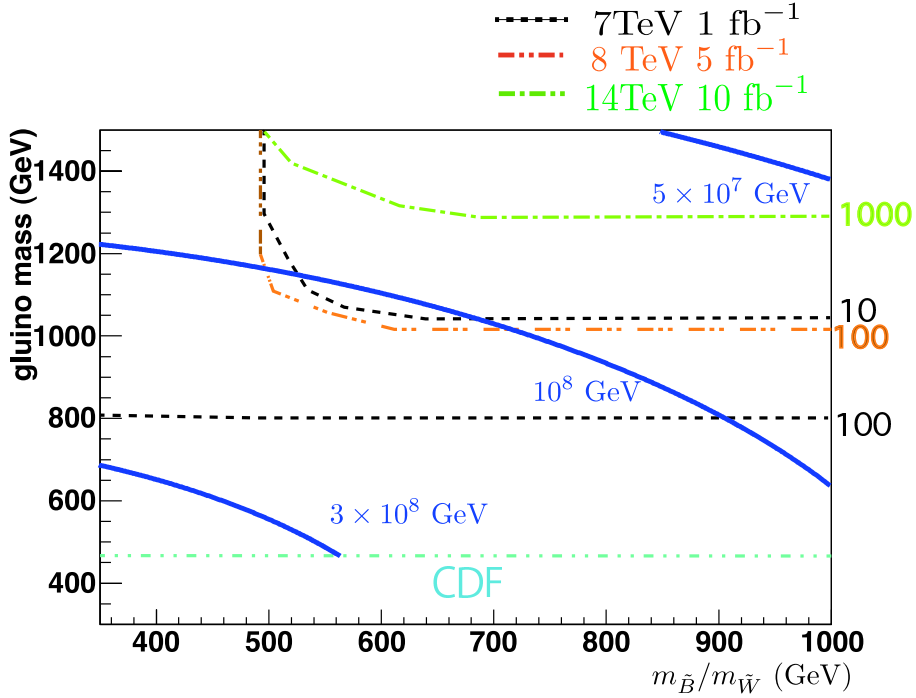}
\end{center}
\caption{The upper bound on the gluino mass for the case (A), where the stau annihilation is dominated by the electroweak processes. The bino and wino masses are varied assuming $m_{\tilde{B}}
=m_{\tilde{W}}$, while the stau mass is fixed as $m_{\stau}=300 {\rm GeV}$.
The solid (blue) lines are upper bounds on the gluino mass.
We also plot the number of stau signatures at LHC with each $\sqrt{s}$ and 
integrated luminosity. 
Here, the events only include the productions of gluinos, charginos and/or neutralinos 
(and staus), and we impose cuts and triggers. The horizontal dashed line around $m_{\tilde{g}}\simeq 470$ GeV
comes from the CDF bound. 
We discuss the collider signature in Section.\ref{collider}.
}
\label{fig:nowino}
\end{figure}
%%%%%%%%%%

This upper bound on the gravitino mass can be easily translated into the 
upper bound on the gluino mass
for a given stau mass and a reheating temperature. 
As we can see from Eq.(\ref{gluinomass}) and Eq.(\ref{Omegah}),
the upper bound on the gluino mass also depends on
the other gaugino masses. In Fig.\ref{fig:nowino}, we show the upper bound on the gluino mass for a given gaugino masses and reheating temperature. 
We set the stau mass to be $300~{\rm GeV}$ in the figure. 
We find the heavier gaugino masses lead to 
the more stringent upper bound on the gluino mass.

\begin{figure}[t]
\begin{center}
\includegraphics[width=12cm]{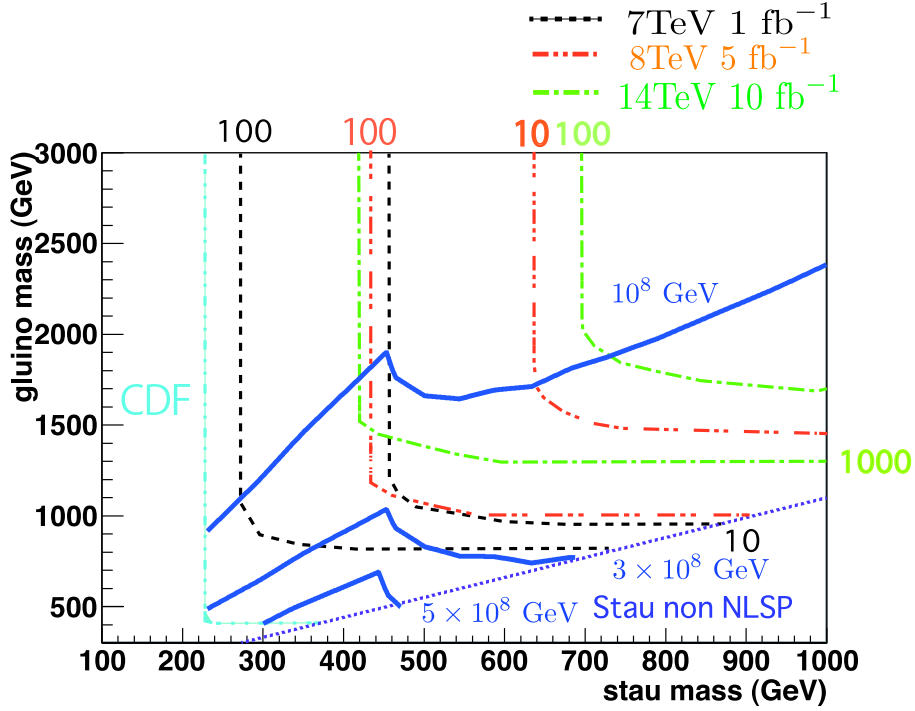}
\end{center}
\caption{Case (A)-(i), where the stau annihilation is dominated by the electroweak interactions, and
$m_{\tilde{B}}=m_{\tilde{W}}=1.1~m_{\stau}$. 
The solid (blue) lines are upper bounds on the gluino mass for various reheating temperatures
$T_{R}$. Contour plots of the number of expected SUSY events at LHC and the
  line of the CDF bound are shown in the same way as Fig.~\ref{fig:nowino} .
  The stau is not the LSP under the dotted (purple) line.}
\label{fig:ew11}
\end{figure}

\begin{figure}[t]
\begin{center}
\includegraphics[width=12cm]{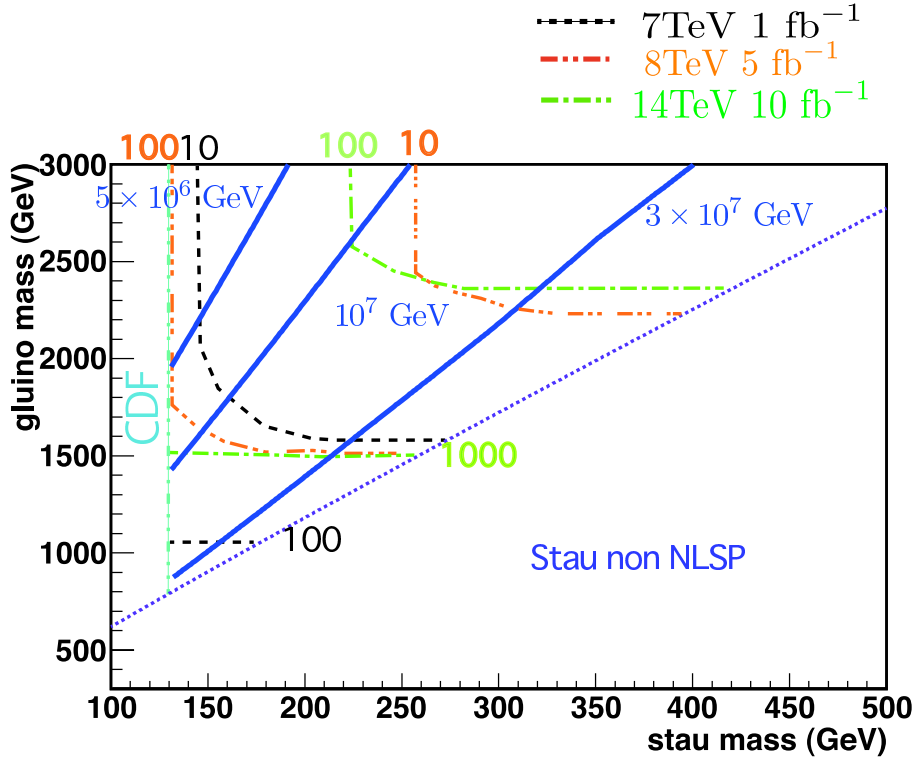}
\end{center}
\caption{Case (A)-(ii), where the stau annihilation is dominated by the electroweak interactions, and
the gaugino masses satisfy the GUT relation. The contour lines are the same as Fig. \ref{fig:ew11}.}
\label{fig:ewgut}
\end{figure}

In the following discussion, we show the upper bounds on the gluino mass for the following two different 
gaugino mass relations:
\begin{enumerate}
\item[(i)]
$m_{\tilde{B}}=m_{\tilde{W}}=1.1m_{\stau}$.
This gives almost the most
 conservative upper bound on the gluino mass (cf. Eq.~(\ref{gluinomass})).
Note that if $m_{\tilde{B}(\tilde{W})}$ is too degenerate with $m_{\stau}$, the
stau abundance is enhanced by the $\tilde{B}(\tilde{W})$ decay, 
resulting in more stringent bounds on $T_R$ (or $m_{\tilde{g}}$) \cite{Asaka:2000zh}.
\item[(ii)]
Gaugino masses satisfy the GUT relation, 
$m_{\tilde{B}}/g_1^2(m_{\tilde{B}})
=m_{\tilde{W}}/g_2^2(m_{\tilde{W}})
=m_{\tilde{g}}/g_3^2(m_{\tilde{g}})$,
where $g_1(\mu), g_2(\mu)$ and $g_3(\mu)$ are the running gauge coupling constants of 
$U(1)_{Y},~SU(2)_L$ and $SU(3)_C$ gauge symmetries at the mass scale $\mu$ respectively. 
 This is the case for the minimal supergravity and the minimal gauge mediation models.
\end{enumerate}
Fig.~\ref{fig:ew11} and  \ref{fig:ewgut} show the results for the cases (i) and (ii), respectively.
{}From the figures, we find the followings;
\begin{itemize}
\item[(i)] $m_{\tilde{B}}=m_{\tilde{W}}=1.1m_{\stau}$ case:
the gluino mass cannot exceed about $2.4$ TeV for $m_{\stau}<1~$TeV
and $T_R\gsim10^8~{\rm GeV}$.
For $T_R>3\ (5)\times 10^8~{\rm GeV}$,
the stau mass is bounded as $m_{\stau}\lsim 700\ (500)~{\rm GeV}$, 
and the gluino mass is bounded as $m_{\tilde{g}}\lsim 1100\ (700)~{\rm GeV}$.
There is no region for $T_R\gsim 7\times 10^8~{\rm GeV}$.
\item[(ii)] The case of the GUT relation:
the constraint on
gluino mass and the reheating temperature are much severer than the case (i).
There is no region for $T_R \gsim 7\times10^7~{\rm GeV}$.
\end{itemize}

%%%%%%%%%%
%%%%%%%%%

\label{sec:ew}
%%%%%%%%%%%%%%%%%%%%%%%%%%
\subsubsection{(B) Stau Annihilation with large stau-stau-light Higgs coupling} 
\label{sec:B}
%%%%%%%%%%%%%%%%%%%%%%%%%%
When the $\stau-\stau-h$ coupling is large, the stau annihilation into the light CP-even Higgs bosons $h$ is enhanced~\cite{enhance1,enhance2}.
The trilinear coupling of the lighter stau and $h$ is given by
\begin{eqnarray}
  {\cal L} = -{\cal A}_{\stau \stau h^0} \stau_1^*\stau_1 h,
  \label{eq:stau-trilinear}
\end{eqnarray}
where the coefficient takes the form of
\begin{eqnarray}
  {\cal A}_{\stau \stau h^0} &\simeq& 
  - \frac{g m_\tau}{2 M_W} (\mu\tan\beta + A_\tau) \sin 2\theta_\tau
  + \frac{g m_\tau^2}{M_W}
  \nonumber \\ &&
  - g_Z M_Z \left[
    \left(-\frac{1}{2}+\sin^2\theta_W\right)\cos^2\theta_\tau
    - \sin^2\theta_W \sin^2\theta_\tau
  \right],
\end{eqnarray}
where $g$, $g_Z$,  $M_W$, $M_Z$, $m_{\tau}$ and $\theta_W$ are the Standard Model parameters, $\mu$ is the Higgsino mass parameter, $\tan \beta$ is the ratio of VEVs
of the two Higgs doublets, and $\theta_{\tau}$ is the mixing angle of the
staus defined by
\begin{eqnarray}
\left(
\begin{array}{c}
\stau_1 \\
\stau_2 \\
\end{array}
\right)
=
\left(
\begin{array}{c c}
\cos \theta_{\tau} &\sin \theta_{\tau}   \\
-\sin \theta_{\tau} &\cos \theta_{\tau} \\
\end{array}
\right)
\left(
\begin{array}{c}
\stau_L \\
\stau_R \\
\end{array}
\right)
.
\end{eqnarray}
This trilinear coupling $ {\cal A}_{\stau \stau h^0} $
  is enhanced when $\mu$, ${\rm tan}\beta$ and 
sin$2\theta_{\tau}$ are large. 
The stau annihilation process via the 
large trilinear coupling may significantly reduce the stau abundance \cite{enhance1,enhance2}. 

\begin{figure}[t]
\begin{center}
\includegraphics[width=10cm]{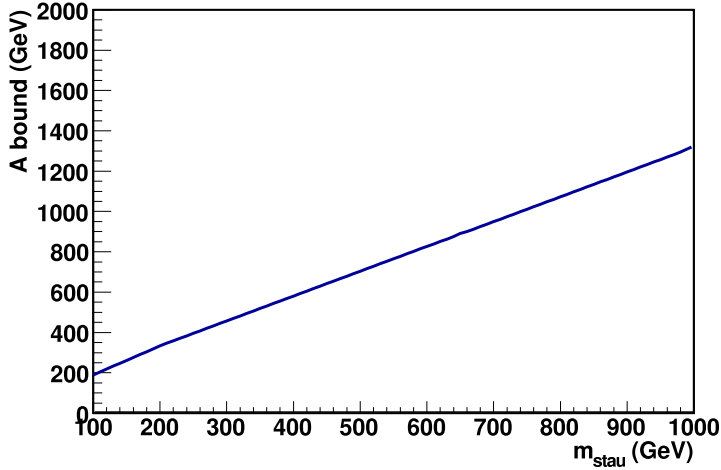}
\caption{CCB bound on the stau-Higgs coupling ${\cal A}_{\stau \stau h^0}$ for $m_h=120$~GeV.
The constraint from the zero temperature decay rate is severer for $m_{\tilde\tau} > 220\,{\rm GeV}$,
while the finite temperature transition dominates the bound for $m_{\tilde\tau} < 220\,{\rm GeV}$.
}
\label{fig:ccbbound}
\end{center}
\end{figure}

However, the large trilinear coupling may create disastrous charge/color breaking (CCB) minima in the $\stau-h$ potential~\cite{enhance1}.
In such a case, the vacuum in our universe is a local minimum, 
and therefore, our vacuum will eventually decay into a global minimum. The condition that the lifetime of our vacuum must be longer than the age of the universe, gives the upper bound 
on the value of ${\cal A}_{\stau \stau h^0}$, as discussed in Appendix \ref{sec:App}.
The resultant upper bound on ${\cal A}_{\stau \stau h^0}$ is shown in Fig.~\ref{fig:ccbbound} for $m_h=120$~GeV.

%%%%%%%%%%%%%%
%%%%%%%%%%%%%%

\begin{figure}[t]
\begin{center}
\includegraphics[width=10cm]{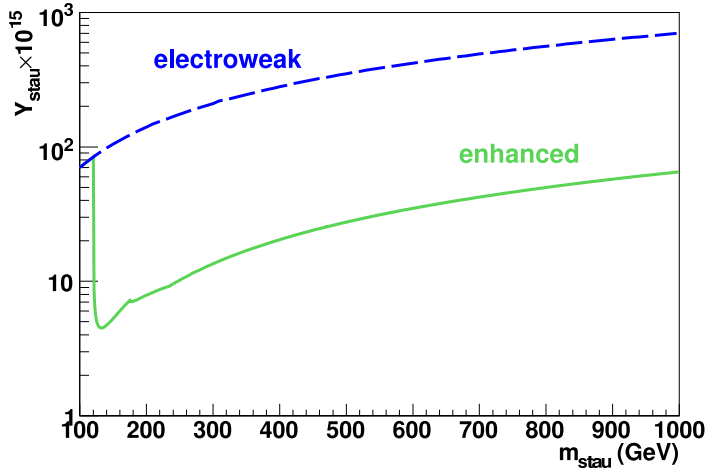}
\caption{Green solid line: $Y_{\stau}$ vs $m_{\stau}$ when $A_{\stau\stau h}$ is the maximal value.
We take $m_h=120~{\rm GeV}$ and $m_t=173~{\rm GeV}$. Blue dashed line: 
$Y_{\stau}$ when the electroweak process is the dominant
stau annihilation process.}
\label{enhanceYt}
\end{center}
\end{figure}

\begin{figure}[t]
\begin{center}
\includegraphics[width=10cm]{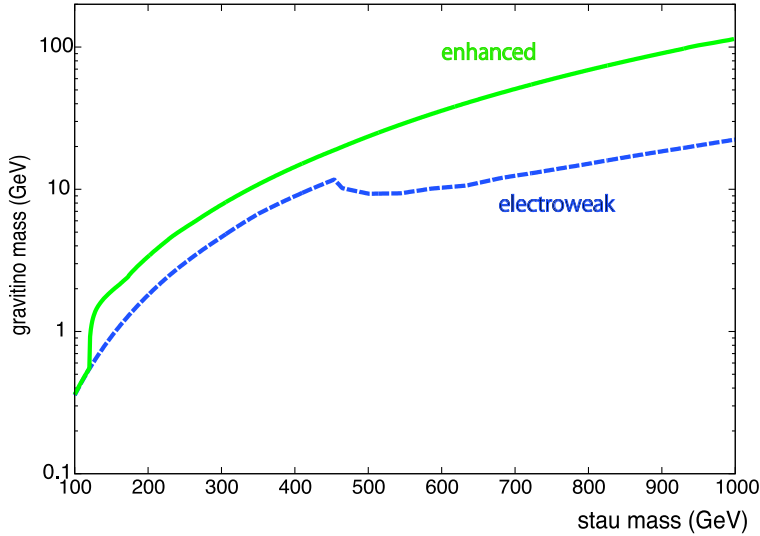}
\end{center}
\caption{Green solid line: gravitino mass bound for the case with an enhanced $h$-stau coupling
. Blue dashed line: gravitino mass bound for the case of the electroweak annihilation.}
\label{fig:gravitino}
\end{figure}

The upper bound on ${\cal A}_{\stau \stau h^0}$ gives the upper bound
on the stau annihilation cross section. Since 
the relic abundance of the stau is given by
\begin{eqnarray}
  Y_{\stau} \simeq 
  1.0\times10^{-15}
  \left(\frac{10^{-5}{\rm GeV^{-2}}}{\langle \sigma v\rangle}\right)
  \left(\frac{200{\rm GeV}}{m_{\stau}}\right),
  \label{eq:stau-abundance}
\end{eqnarray}
the upper bound on the stau annihilation cross section gives the lower bound on 
$Y_{\stau}$,
where $\sigma$ is the annihilation cross section of stau, $v$ is the relative velocity of staus and
$\langle \rangle$ denotes the thermal average. In the large trilinear coupling case,
$\langle \sigma v\rangle$ takes the form of
\begin{eqnarray}
\langle \sigma v\rangle \simeq 
  \frac{{\cal A}_{\stau \stau h^0}^4}{64\pi m^6_{\stau}} f_h +
  \frac{3Y_t^2{\cal A}_{\stau \stau h^0}^2}{128\pi m_{\stau}^4} f_t,
  \label{eq:cross-section}
\end{eqnarray}
where
\begin{eqnarray} 
f_h = \frac{\sqrt{1-r_h}}{(1-r_h/4)^2} \theta(1 - r_h),~~~
f_t = \frac{(1-r_t/2)\sqrt{1-r_t}}{(1-r_h/4)^2} \theta(1 - r_t),
\end{eqnarray}
with $r_h = m_h^2/m_{\tilde \tau}^2$ and $r_t = m_t^2/m_{\tilde \tau}^2$. 
If $m_{\stau}$ is larger than $m_h$, the annihilation 
into Higgs bosons via the large trilinear coupling is possible,
and the first term in Eq.(\ref{eq:cross-section}) becomes nonzero.
In the same way, the second term in Eq.(\ref{eq:cross-section})
becomes nonzero when $m_{\stau}$ is larger than $m_t$.
In the following, we take $m_h=120~{\rm GeV}$ and $m_t=173~{\rm GeV}$.
 The lower bound on $Y_{\stau}$ in 
 this large ${\cal A}_{\stau \stau h^0}$ scenario
 is shown in Fig.\ref{enhanceYt}. 
 For comparison, 
 we also show the $Y_{\stau}$ when the electroweak process is the dominant
stau annihilation process in the same figure.
 We find that $Y_{\stau}$ in the large ${\cal A}_{\stau \stau h^0}$ scenario
 is significantly smaller than that of the electroweak annihilation,  but it
  cannot be smaller than $10^{-15}$.
Therefore,
 the bound from the $^6{\rm Li}$ overproduction by the catalyzed effect severely constrains the gravitino mass \cite{bbn2},
\begin{eqnarray}
  m_{3/2} \lsim \begin{array}{ll}
  0.4 {\rm GeV}-115 \rm{GeV}& (100{\rm GeV}<m_{\stau}<1000 {\rm GeV})  
\end{array} 
\end{eqnarray}
as shown in Fig.\ref{fig:gravitino}.
  
%%%%%%%%%%%%%%  
%%%%%%%%%%%%%%
%%%%%%%%%%%%%%

\begin{figure}[t]
\begin{center}
\includegraphics[width=12cm]{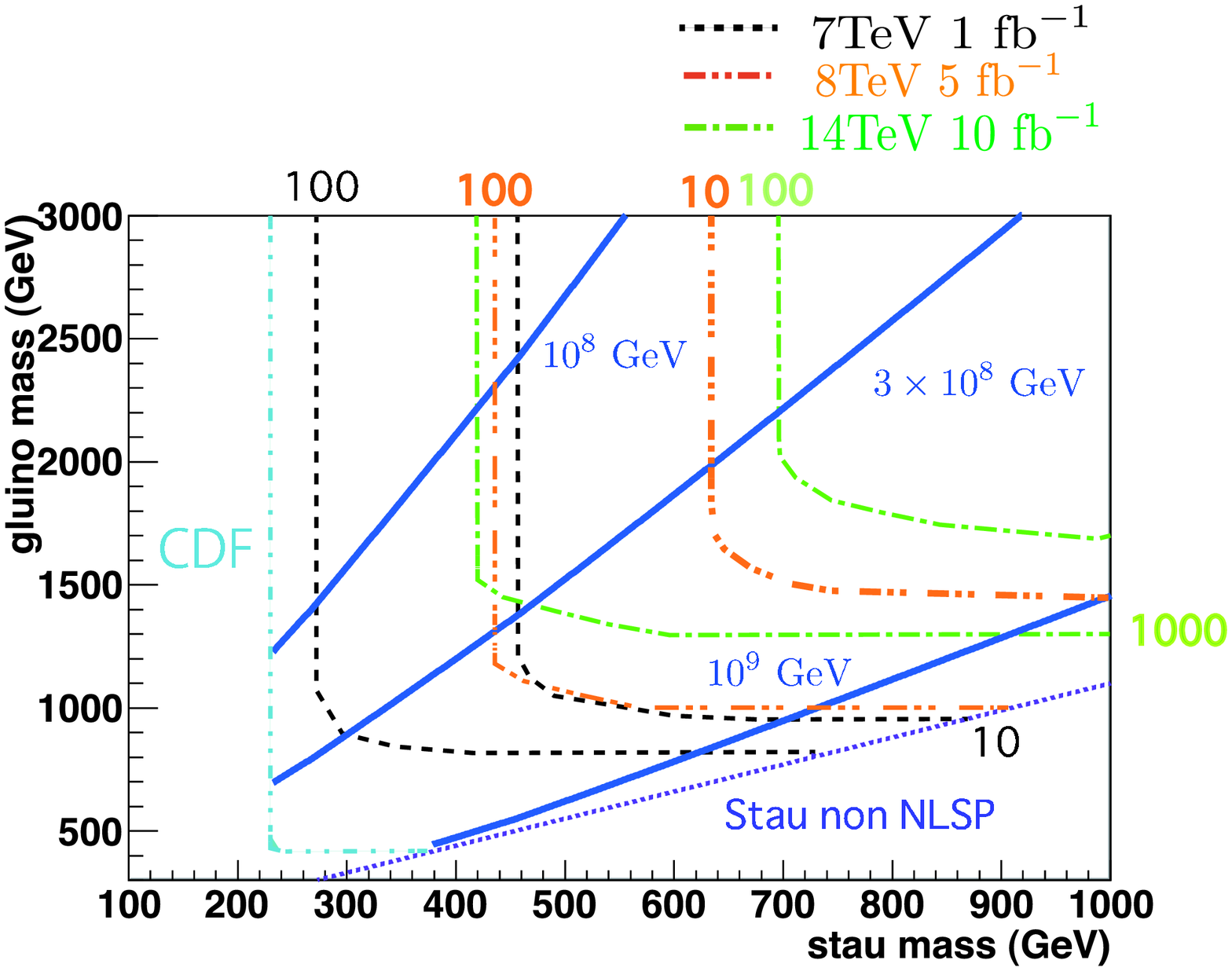}
\end{center}
\caption{
Case (B)-(i), where the stau annihilation is enhanced by the $h$-stau-coupling, and
$m_{\tilde{B}}=m_{\tilde{W}}=1.1~m_{\stau}$. The contour lines are the same as Fig. \ref{fig:ew11}.}
\label{fig:11}
\end{figure}

\begin{figure}[t]
\begin{center}
\includegraphics[width=12cm]{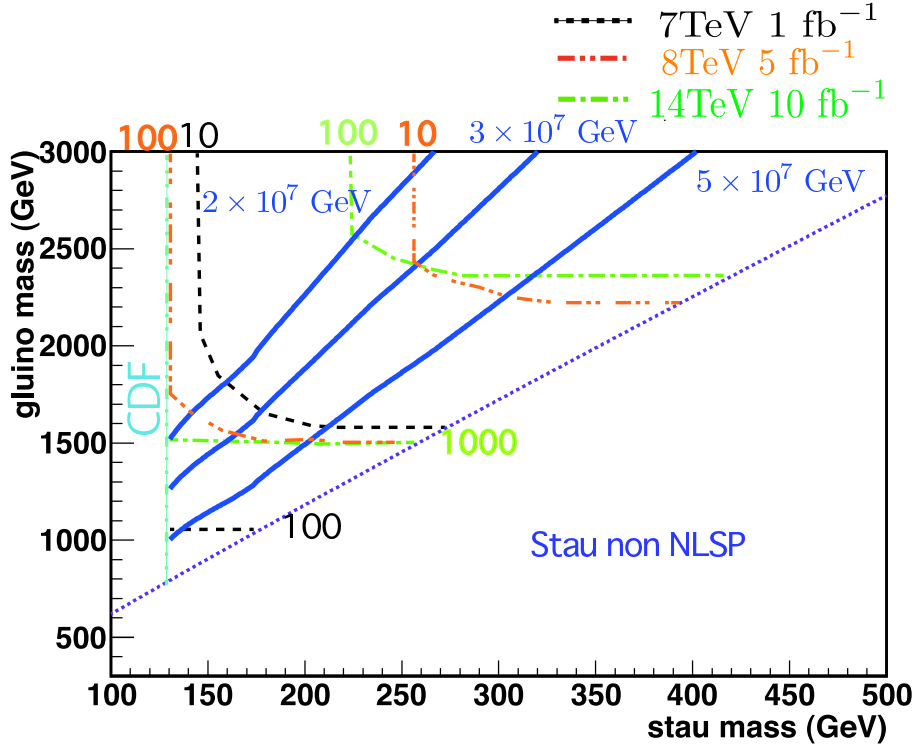}
\end{center}
\caption{
Case (B)-(ii), where the stau annihilation is enhanced by the $h$-stau-coupling, and
the gaugino masses satisfy the GUT relation. The contour lines are the same as Fig. \ref{fig:ew11}.}
\label{fig:gut}
\end{figure}

This gravitino mass upper bound leads to the upper bound on the 
gluino mass for a given stau mass
and reheating temperature. 
As in the case in Section \ref{sec:ew}, we show the upper bounds on the gluino mass
with the two gaugino mass relations:(i) $m_{\tilde{B}}=m_{\tilde{W}}=1.1m_{\stau}$, and
(ii) the GUT relation. The upper bound with the relation (i) is shown in Fig.\ref{fig:11},
 while that with the relation (ii) is shown in Fig.\ref{fig:gut}. 
 From Fig.\ref{fig:11} and Fig.\ref{fig:gut}, we find the followings;
\begin{enumerate}
\item[(i)] $m_{\tilde{B}}=m_{\tilde{W}}=1.1m_{\stau}$ case:
the behavior of the gluino mass upper bound in Fig.\ref{fig:11} are similar to that in
Fig.\ref{fig:ew11} for
$m_{\stau}\le 450~{\rm GeV}$, since the gravitino mass upper bound is determined by
the constraints from $^6{\rm Li}$ in the region. On the other hand, 
the behavior is
quite different
for $m_{\stau}>450~{\rm GeV}$, because the gravitino mass upper bound 
in the case (A) is determined by the constraints from the deuterium, while the
upper bound in the case (B) is still determined by 
the constraints from $^6{\rm Li}$. It is also remarkable that 
$T_R>10^{9}~{\rm GeV}$ is possible.

\item[(ii)] The case of the GUT relation:
the behavior of the gluino mass upper bound is almost the same as Fig.\ref{fig:ewgut}.
We also see that the
behavior of the gluino mass bound slightly 
changes at $m_{\stau}=m_t$, because the
annihilation into top quark becomes possible for $m_{\stau}>m_t$.
\end{enumerate}

%%%%%%%%%%
%%%%%%%%%%

%%%%%%%%%%
%%%%%%%%%%

%%%%%%%%%%%%%%%%%%%%%%%%%%
\subsubsection{(C) Stau Annihilation near the pole of the Heavy Higgs Boson}
\label{sec:C}
%%%%%%%%%%%%%%%%%%%%%%%%%%
The stau annihilation cross section can be considerably enhanced in case that $m_{H}\simeq 2 m_{\stau}$
and $A_{\tau}$, $\tan\beta$ and $\sin 2\theta_{\tau}$ are large \cite{enhance2}. The trilinear coupling of the lightest 
stau with $H$ takes the form of
\begin{eqnarray}
{\cal L}=-{\cal A}_{\stau\stau H}\stau^{\ast}_1 \stau_1 H,
\end{eqnarray}
where the coefficient is given by
\begin{eqnarray}
{\cal A}_{\stau\stau H}&\simeq&-\frac{gm_{\tau}}{2M_W}(A_{\tau}\tan\beta-\mu)\sin 2\theta_{\tau}
+\frac{gm^2_{\tau}}{M_W}\tan\beta.
\end{eqnarray} 
in the large ${\rm tan}\beta$ region.
When $A_{\tau}\tan\beta$ is large, the first term in the right hand side becomes enhanced.
In such large  $A_{\tau}\tan\beta$ case, the annihilation through $H$ exchange can be the dominant process, and if $m_{H}$ satisfies the resonance condition 
 $m_{H}\simeq 2 m_{\stau}$, it is enhanced greatly. 

\begin{figure}[t]
\begin{center}
\includegraphics[width=10cm]{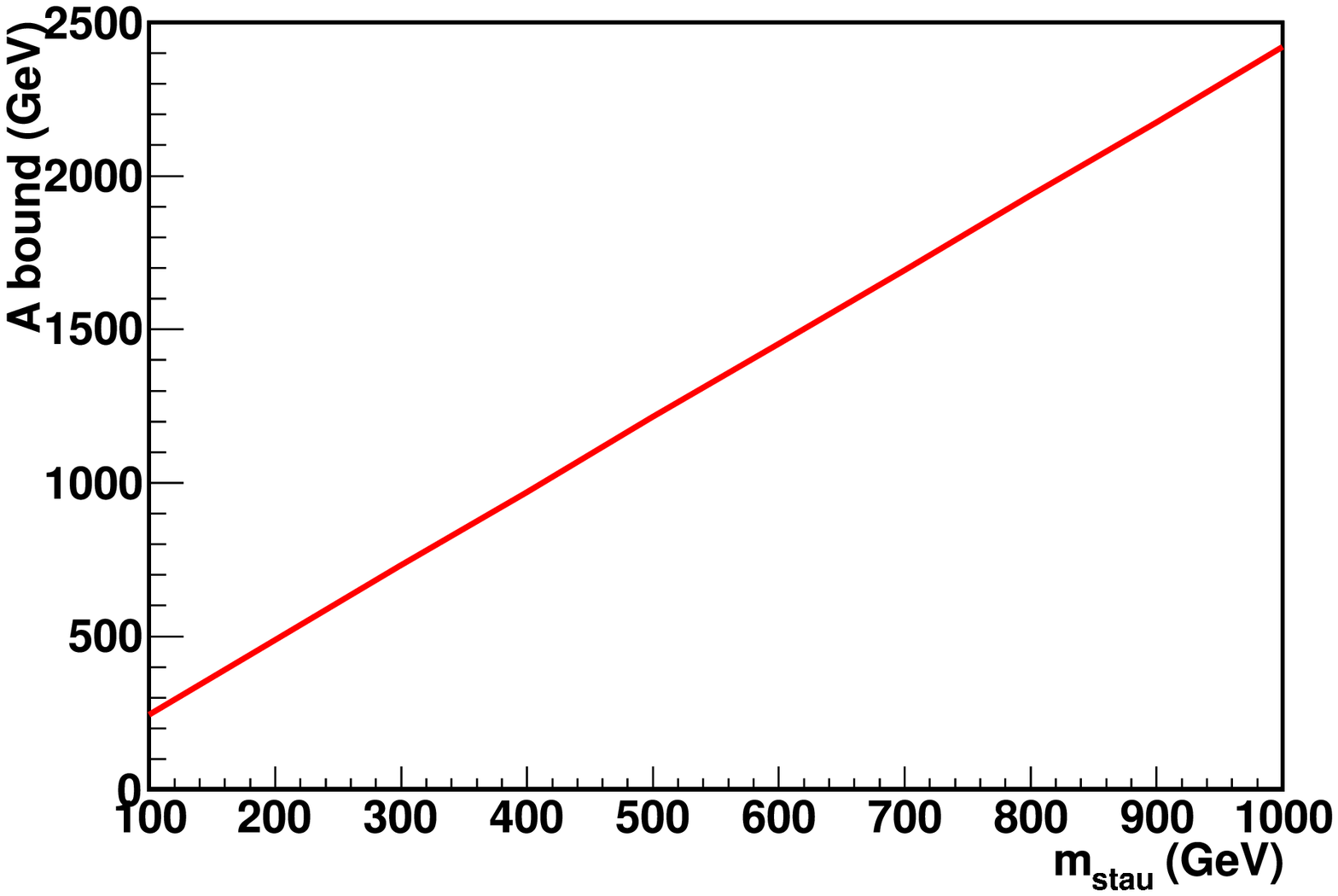}
\caption{CCB bound on the stau-Higgs coupling ${\cal A}_{\stau \stau H}$. The finite temperature transition dominates the bound for $m_{\tilde\tau} > 100\,{\rm GeV}$.}
\label{fig:CCBlarge}
\end{center}
\end{figure}

The large ${\cal A}_{\stau\stau H}$ may generate the disastrous CCB minimum, 
and therefore,
the value of ${\cal A}_{\stau\stau H}$ is constrained as in the case of the enhanced $h$-stau trilinear coupling. We obtain the upper bound on ${\cal A}_{\stau\stau H}$ as we
discuss in Appendix \ref{sec:App}. The bound on ${\cal A}_{\stau\stau H}$ is shown in Fig.~\ref{fig:CCBlarge} for $m_H = 2m_{\stau}$.

%%%%%%%%%%%%%%
%%%%%%%%%%%%%%
\begin{figure}[t]
\begin{center}
\includegraphics[width=10cm]{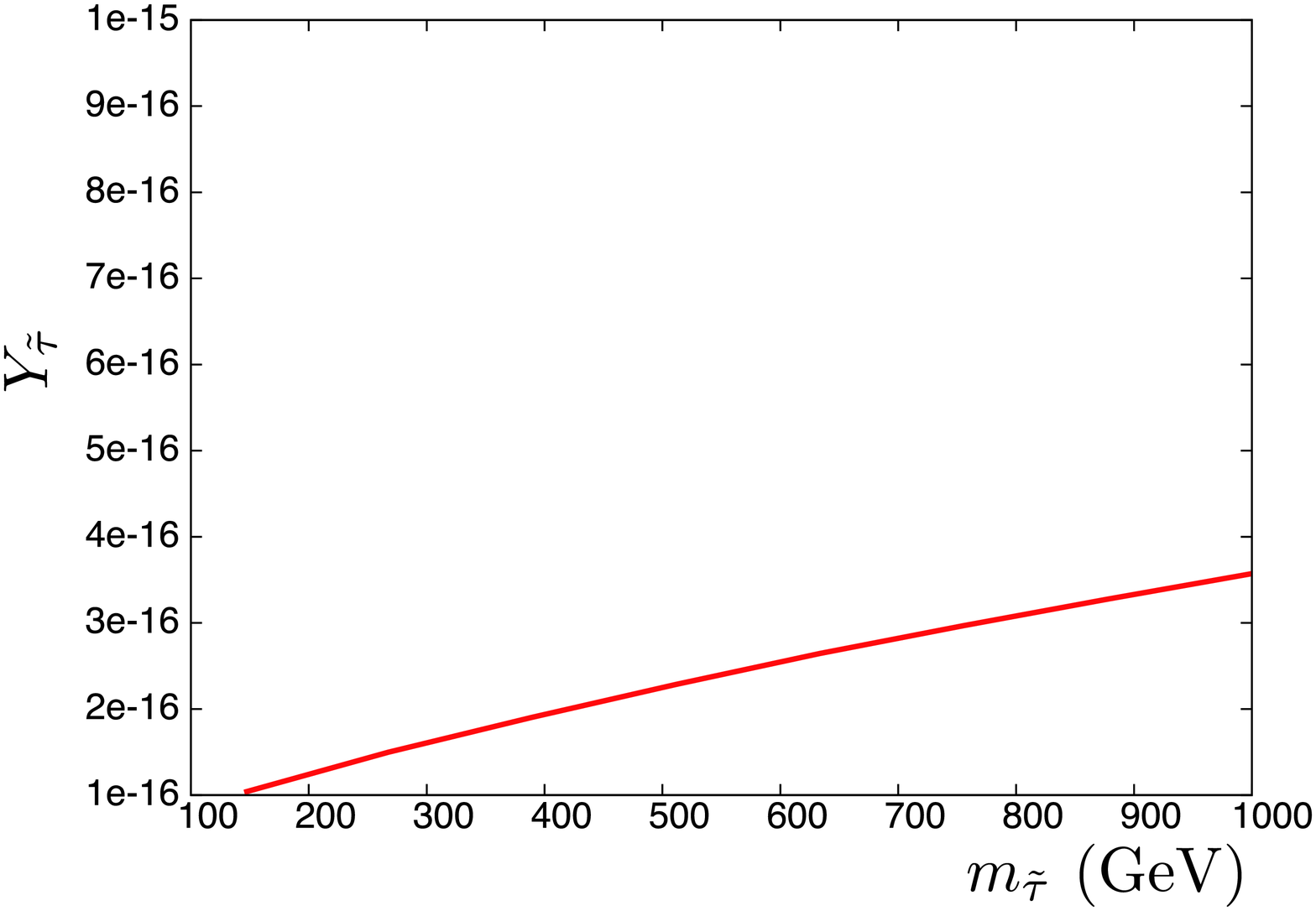}
\caption{$Y_{\stau}$ vs $m_{\stau}$ when ${\cal A}_{\stau\stau H}$=$1.3\times m_{\stau}$, and
$m_H=2  m_{\stau}$.}
\label{Ytheavy}
\end{center}
\end{figure}
%%%%%%%%%%%%%%

In this case, the stau relic abundance can be sufficiently reduced to 
avoid the BBN constraint, i.e., 
$Y_{\stau}<10^{-15}$.
For illustration, we take $m_H=2m_{\stau}$
and ${\cal A}_{\stau\stau H}$=$1.3\times m_{\stau}$,
which is below the 
upper bound on ${\cal A}_{\stau\stau H}$
in $100{\rm GeV}<m_{\stau}<1000{\rm GeV}$, 
as can be seen from Fig. \ref{fig:CCBlarge}.
The resultant abundance
$Y_{\stau}$ is shown in Fig.~\ref{Ytheavy}, where
we use the program micrOMEGA2.4 \cite{micromega} to calculate the relic abundance.
We see that $Y_{\stau}$ can be less than $10^{-15}$ for 
$100{\rm GeV}<m_{\stau}<1000{\rm GeV}$, and therefore
BBN does not provide any constraints on the gravitino mass.

There is another constraint from
the CMB spectrum distortion~\cite{CMBdistort}.
 However,
the difference between the stau mass and
the gravitino mass upper bound $m_{3/2}^{max}$  is 
small,
$(m_{\stau}-m^{max}_{3/2}(m_{\stau}))/m_{\stau}\lsim 0.01$.
%%%%%%%%%
Thus, we take  $m^{max}_{3/2}(m_{\stau})=m_{\stau}$, for simplicity.

\begin{figure}[t]
\begin{center}
\includegraphics[width=12cm]{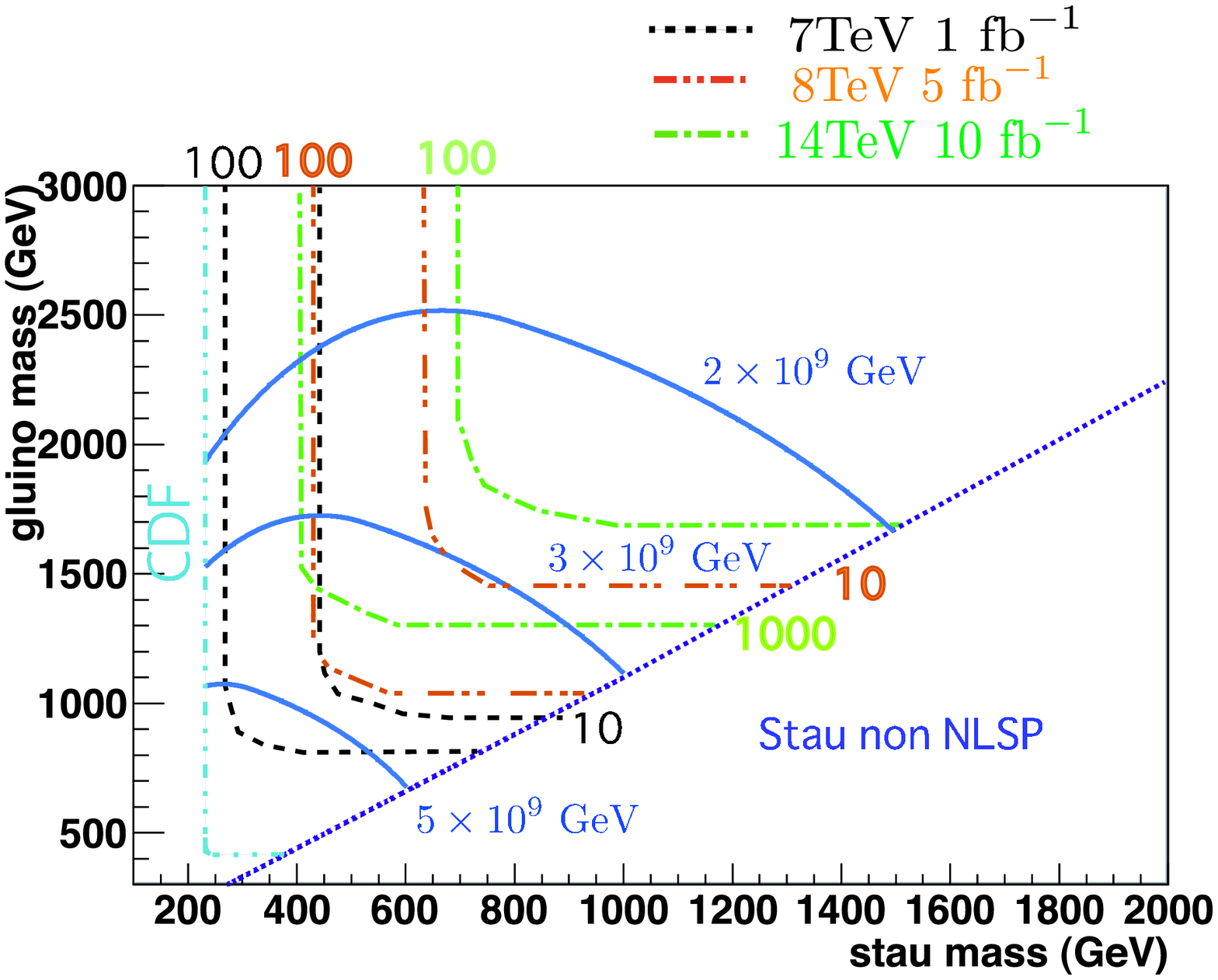}
\end{center}
\caption{
Case (C)-(i), where $m_{H}\simeq 2m_{\stau}$ and the $H$-stau-coupling is enhanced, and
$m_{\tilde{B}}=m_{\tilde{W}}=1.1~m_{\stau}$. The contour lines are the same as Fig. \ref{fig:ew11}.}
\label{fig:resonance11}
\end{figure}

\begin{figure}[t]
\begin{center}
\includegraphics[width=12cm]{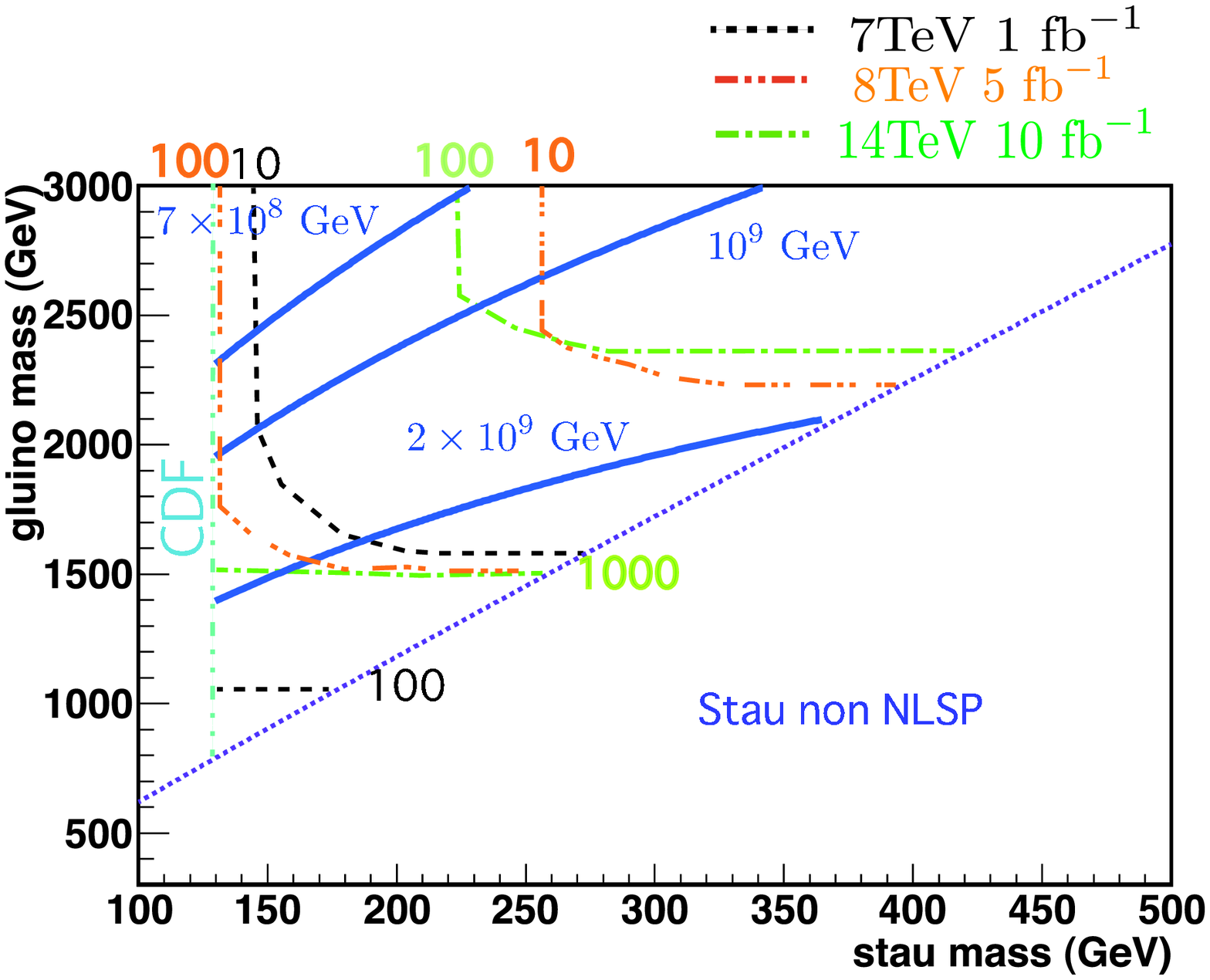}
\end{center}
\caption{
Case (C)-(ii), where $m_{H}\simeq 2m_{\stau}$ and  the $H$-stau-coupling is enhanced, and
the gaugino masses satisfy the GUT relation. The contour lines are the same as Fig. \ref{fig:ew11}.}
\label{fig:resonancegut}
\end{figure}

The upper bound on the gravitino mass leads to the
upper bound on the gluino mass for a given stau mass and 
reheating temperature. As in the case in Section \ref{sec:ew} and Section \ref{sec:B},
we show the gluino mass upper bound
(i) for $m_{\tilde{B}}=m_{\tilde{W}}=1.1\times m_{\stau}$ in Fig.~\ref{fig:resonance11}, 
and (ii) when there is the GUT relation between gaugino masses in
Fig.~\ref{fig:resonancegut}.
Note that, in the larger stau mass region, 
the gluino mass is maximized 
when the gravitino mass is smaller than $m^{max}_{3/2}(m_{\stau})$.
That is because the upper bound on the gluino mass drops down
in the higher gravitino mass region as shown in Fig.\ref{fig:gralsp}. From 
Fig.~\ref{fig:resonance11}
and
Fig.~\ref{fig:resonancegut},
we find followings;
\begin{enumerate}
\item[(i)] $m_{\tilde{B}}=m_{\tilde{W}}=1.1\times m_{\stau}$ case:
Interestingly, 
$T_R>2\times10^{9}~{\rm GeV}$ which is required by thermal leptogenesis
is possible in the region $m_{\stau} \lsim 1500~{\rm GeV}$.
We can even achieve $T_R>5\times10^{9}~{\rm GeV}$.
The gluino mass bound becomes lower
in the larger 
$m_{\stau}$ region because
the Bino mass and the Wino mass becomes heavier for large stau mass, and the constraints
of the gravitino overclosure becomes severer. 
\item[(ii)]When there is the GUT relation:
$T_R>2\times10^{9}~{\rm GeV}$ is possible in $m_{\stau} \lsim 350~{\rm GeV}$.
The gluino mass is bounded as $m_{\tilde{g}}\lsim 2~{\rm TeV}$ when
$T_R>2\times10^{9}~{\rm GeV}$.
\end{enumerate}

%%%%%%%%%%%%%%
%%%%%%%%%%%%%%
%%%%%%%%%%%%%%

%%%%%%%%%%%%%%%%%%%%%%%%%%%
\section{Collider Signatures}
%%%%%%%%%%%%%%%%%%%%%%%%%%%
\label{collider}
Let us study the collider detectability of the scenario of the high reheating temperature. 
In the previous section, it was shown that the gluino mass is constrained to be less than a few TeV 
to realize a high reheating temperature.
 Light SUSY particles have been already excluded 
by direct searches in Tevatron, while TeV colored SUSY particles are in the reach of sensitivity of LHC. 
If heavy SUSY particles are produced at a collision, they subsequently decays into lighter SUSY 
particles with radiating SM particles, and promptly generate the NLSP stau in the end of the decay 
chain. Since the gravitino mass is as large as ${\cal O}(1-100)$GeV in the high 
reheating temperature scenario, the NLSP stau is long-lived enough to be observed as a stable 
particle in the detectors. Noting that the stau has an electromagnetic charge, we expect the events with 
charged tracks when SUSY particles are produced at collisions. In this section, we discuss the Tevatron 
bound and the LHC sensitivity of the high reheating temperature scenario. 

Before proceeding to the collider study, we summarize the tools for the numerical analysis. 
We use PYTHIA 6.4.22 \cite{pythia} to study the kinematics and to estimate the cross sections except for 
those of the colored SUSY particle productions. The gluino and squark production cross sections are 
estimated by the program Prospino2 \cite{prospino} at the NLO level. The Tevatron and LHC detectors 
are simulated by the package PGS4 \cite{pgs}.
%%%%%%%%%%%%%%%%%%%%%%%%%%
\subsection{Tevatron Bound}
%%%%%%%%%%%%%%%%%%%%%%%%%%
At the Tevatron experiments, the stau is expected to behave as a heavy muon, namely a charged 
massive and long-lived particle. As long as the stau has a large velocity, it is not distinguishable 
from the muon. When the transverse momentum, $p_T$, of the muon is large, it has 
a large velocity of $\beta \simeq 1$, while the stau is likely to have a lower speed. Such a high-$p_T$ and 
low speed ``muon" has been searched for in Tevatron by measuring the time of flight 
\cite{tevatron,Abazov:2008qu}. 
According to \cite{tevatron}, the events are selected by the following trigger:
\begin{itemize}
\item the highest $p_T$ ``muon" candidate has $p_T$ larger than 20 GeV which 
satisfies an isolation condition $E_T (0.4)/p_T (\mu) < 0.1$,
 \end{itemize}
where $E_T (0.4)$ is the sum of the transverse energy within a cone 
$R =0.4$ around the candidate, excluding the energy deposited by the muon candidate
itself, and $p_{T}(\mu)$ is the transverse momentum of the highest 
$p_T$ muon candidate. Note that the long-lived charged massive particle 
is identified as a
``muon" candidate.
The events that satisfy the above trigger condition are read out.
The Tevatron constraint is that 
the production cross section of the long-lived charged massive particle 
which runs toward the direction $|\eta|<0.7$
with $p_T>40 {\rm GeV}$, and with
the velocity $0.4<\beta<0.9$, is smaller than $10~{\rm fb}$ at the 95\% C.L..
The constraint  gives the lower bound on masses of SUSY particles.

We show the constraints from Tevatron in Figs.\ref{fig:nowino}--\ref{fig:ewgut},
 Fig.\ref{fig:11}, Fig.\ref{fig:gut}, Fig.\ref{fig:resonance11} and Fig.\ref{fig:resonancegut}. 
The constraints in Fig.\ref{fig:ew11}, Fig.\ref{fig:11}, and Fig.\ref{fig:resonance11} are the same,
which corresponding to the cases (A)-(i), (B)-(i), and (C)-(i), respectively.
Similarly, the constraints in Fig.\ref{fig:ewgut}, Fig.\ref{fig:gut}, and Fig.\ref{fig:resonancegut} 
(corresponding to the cases (A)-(ii), (B)-(ii), and (C)-(ii), respectively) 
are the same.  
 We only consider productions of the gauginos and the lighter staus, 
 and do not consider the production of the other
 scalar particles. This is realized
 when their masses are relatively heavy. 
 In the case that the gluino mass is relatively small, gluino pair production is the main production channel. That is the case in the regions around the horizontal lines in Fig.\ref{fig:nowino},
 Fig.\ref{fig:ew11}, Fig.\ref{fig:11} and Fig.\ref{fig:resonance11}.
When the gluino mass is large, the stau direct production, chargino-neutralino
and chargino-chargino pair productions are the main production channel.
The vertical lines of the CDF constraints in Fig.\ref{fig:ew11} 
are determined by the chargino-neutralino
and chargino-chargino pair production cross sections.
The same situation holds in the regions around
 the vertical lines in Fig.\ref{fig:11} and Fig.\ref{fig:resonance11}.
On the other hand, the vertical line
in Fig.\ref{fig:ewgut} is determined by
the stau direct production. That is also the case in the regions around the vertical lines in
Fig.\ref{fig:gut} and Fig.\ref{fig:resonancegut}.

%%%%%%%%%%%%%%%%%%%%%%%%%%
\subsection{LHC signatures}
%%%%%%%%%%%%%%%%%%%%%%%%%%
The scenarios of the high reheating temperature predict the gluino mass being less than a few 
TeV. Although the Tevatron energy is not large enough to cover the mass range, the LHC is suited 
for detecting the particle. The LHC is running at the center-of-mass energy $\sqrt{s}=7~{\rm 
TeV}$ and the integrated luminosity is planned to become up to a few ${{\rm fb}^{-1}}$ in 2011. The schedule 
of 2012 is still in discussion: an optimistic scenario is $\sqrt{s}=8~{\rm TeV}$
 with $L_i \sim 10{\rm fb}^{-1}$. After the upgrade, the collider is aimed to run at $\sqrt{s}=14~{\rm TeV}$. 
In the following, we discuss the LHC sensitivity of the high reheating temperature scenario in the 
three setups: $\sqrt{s}=7~{\rm TeV}$ with $L_i=1~{\rm fb}^{-1}$, $\sqrt{s}=8~{\rm TeV}$ with 
$L_i=5~{\rm fb}^{-1}$ and  $\sqrt{s}=14~{\rm TeV}$ with $L_i=10~{\rm fb}^{-1}$.

Analogous to the Tevatron, the stau signature is the charged track of the low speed muon-like 
particle with high $p_T$ within the detectors. First of all, the SUSY events are selected by imposing 
a trigger menu, and then analyzed off-line with cut conditions in order to be distinguished from the 
background. In the following analysis, we impose the following trigger conditions,
\begin{itemize}
\item at least one isolated electron has $p_{\rm T}>20$~GeV,
\item at least one isolated muon has $p_{\rm T}>40$~GeV,
\item at least one isolated tau has $p_{\rm T}>100$~GeV,
\item at least one jet has $p_{\rm T}>200$~GeV,
\item at least three jet has $p_{\rm T}>100$~GeV,
\item at least one isolated stau has $p_{\rm T}>40$~GeV within the bunch,
\item at least two staus have $p_{\rm T}>40$~GeV within the bunch.
\end{itemize}
If any one of these conditions is satisfied, the event is read out.
In our simulation, the isolation conditions on the electron and the tau relies on PGS4, while those of the 
muon and the stau are
\begin{enumerate}
\item the summed $p_{\rm T}$ in a $R = 0.4$ cone around 
the particle (excluding the particle itself) is less than $5~{\rm GeV} $,
\item the ratio of $E_{\rm T}$ in a $3\times3$ calorimeter array around the particle (including the 
particle's cell) to $p_T$ of the particle is less than $0.1125$.
\end{enumerate}
The bunch condition, i.e., the condition that the stau has a velocity large enough to reach the muon 
trigger detector before the next bunch collides, is necessary for the stau triggers, since otherwise they 
do not work correctly \cite{reconstruction}. We require $\beta>0.7$ for the stau propagating in the 
barrel region ($|\eta|<1.0$) and $\beta>0.8$ in the endcap region ($1.0<|\eta|<2.8$) \cite{Aad:2009wy}.

Among the events that are read out by the triggers, the SUSY signals are required to satisfy the 
following cut conditions,
\begin{itemize}
\item $p_{{\rm T}}>20~{\rm GeV}$
\item $0.5<\beta<0.9$
\item $|\eta|<2.5$.
\end{itemize}
Although the muon productions are the relevant standard model backgrounds of the stau signals, 
they are significantly reduced to be almost zero by especially the first two cuts, since the high 
$p_{T}$ muons have $\beta\simeq 1$. The last condition is added because the stau is detected in 
the muon detector \cite{Aad:2009wy,reconstruction}.

The number of events is reduced to roughly $50 \%\sim 90\%$ by the 
above trigger conditions.
When the stau direct production or
chargino/neutralino pair production is the dominant production channel,
roughly a half of the triggered events are selected by the stau and the others are selected
by jets and taus. On the other hand, when the gluino production is dominant, 
most of the triggered events are selected by jets and taus. 
Note that we put the trigger efficiency as unity for simplicity.
In the real detector system, however,
the efficiency for single jet and for single tau is not good. 
 Thus the number of the events selected only by single jet or single tau
 (roughly $10\%\sim 40\%$ of the triggered events) may be further reduced
 if we consider the trigger efficiency.
 %%%%%%%%%%%%%%%% 

We also have to consider the reconstruction efficiency
of the stau. The reconstruction efficiency varies from 0.1 to 0.9 for $\beta \simeq 0.5\sim 0.9$
according to ATLAS CSC studies \cite{Aad:2009wy}.
The analysis can be improved by the method studied in \cite{reconstruction}, which provides 
the efficiency more than 90\% for $\beta \gsim 0.5$. In the following, we assume the 
efficiency for the stau with $\beta > 0.5$ to be
100\% and for the stau with $\beta \leq 0.5$ to be zero for simplicity.

In Fig.\ref{fig:nowino}--\ref{fig:ewgut}, Fig.\ref{fig:11}, Fig.\ref{fig:gut},
Fig.\ref{fig:resonance11} and Fig.\ref{fig:resonancegut}, we show the number of 
stau signals that satisfy the trigger conditions and the cut conditions 
at LHC for $\sqrt{s}=7~{\rm TeV}$ with $L_i=1~{\rm fb}^{-1}$,
$\sqrt{s}=8~{\rm TeV}$ with $L_i=5~{\rm fb}^{-1}$, 
 and $\sqrt{s}=14~{\rm TeV}$ with $L_i=10~{\rm fb}^{-1}$. 
 The contour lines in Fig.\ref{fig:ew11}, Fig.\ref{fig:11}, and Fig.\ref{fig:resonance11} are the same,
which corresponding to the cases (A)-(i), (B)-(i), and (C)-(i), respectively.
Likewise, the contour lines in Fig.\ref{fig:ewgut}, Fig.\ref{fig:gut}, and Fig.\ref{fig:resonancegut} 
(corresponding to the cases (A)-(ii), (B)-(ii), and (C)-(ii), respectively) 
are identical.  
We only consider gaugino production and stau direct production at the collision 
 as the production channels, and do not consider the production of the
 other scalar particles.
As in the case of the Tevatron constraints, the
gluino pair production is the main  channel
if the gluino mass is relatively small.
 That is the case in the regions around the horizontal lines in Fig.\ref{fig:nowino},
 Fig.\ref{fig:ew11}, Fig.\ref{fig:11} and Fig.\ref{fig:resonance11}.
When the gluino mass is large, the stau direct production, chargino-neutralino
and chargino-chargino pair productions are the main production channel.
The vertical lines of the LHC signatures in Fig.\ref{fig:nowino}
are determined by the chargino-neutralino
and chargino-chargino pair production cross sections.
The same situation holds in the regions around
 the vertical lines in  Fig.\ref{fig:ew11}, Fig.\ref{fig:11}, Fig.\ref{fig:resonance11},
 and the horizontal lines in 
Fig.\ref{fig:ewgut}, Fig.\ref{fig:gut}, Fig.\ref{fig:resonancegut}.
On the other hand, the vertical line
in Fig.\ref{fig:ewgut} is determined by
the stau direct production. That is also the case in the regions around the vertical lines in
Fig.\ref{fig:gut} and Fig.\ref{fig:resonancegut}.

Now let us discuss the  implications of these LHC signatures 
on the reheating temperature. 
Combining the cosmological upper bound on the gluino mass discussed in Sec.~\ref{cosmology} with the results obtained in this section, 
we can see the minimal number of stau signals 
in each reheating temperature. 
For example, if
 $T_R$ is larger than $3\times10^{8}~{\rm GeV}$  and the
 main stau annihilation process is the electroweak process, 
 more than 10 staus are observed for $\sqrt{s}=7~{\rm TeV}$ with $L_i=1~{\rm fb}^{-1}$. 
 Also, more than 10 staus are observed 
for $\sqrt{s}=14~{\rm TeV}$ and $L_i=10~{\rm fb}^{-1}$,  
 if $T_R$ is larger than $10^{8}~{\rm GeV}$, $m_{\stau}<1000~{\rm GeV}$, and 
the electroweak process is dominant.
In TABLE \ref{result},
 we summarize the range of the reheating temperature with which more than 10 stau signals are expected in
each scenario, $\sqrt{s}$, and $L_i$.   It is remarkable that $T_{R}\gsim2\times10^{9}~{\rm GeV}$
 which is required by the thermal leptogenesis is all covered at 
 $\sqrt{s}=14{\rm TeV}$ and $L_i=10~{\rm fb}^{-1}$.

 \begin{table}[t]
\begin{center}
\begin{tabular}{|c|c|c|c|} 
\hline
Annihilation Process&Gaugino Masses&$\sqrt{s}$, $L_i$&Covered reheating temperature \\ \hline
\hline
(A) electroweak 
&(i) $m_{\tilde{B}}=m_{\tilde{W}}=1.1m_{\stau}$ 
&$7{\rm TeV}$,~$1{\rm fb}^{-1}$&$T_{R}\gsim3\times 10^{8}~{\rm GeV}$
\\ \cline{3-4}
&(Fig.~\ref{fig:ew11})&$8{\rm TeV}$,~$5{\rm fb}^{-1}$&
$T_{R}\gsim10^{8}~{\rm GeV}$
for $m_{\stau}\lsim600{\rm GeV}$
\\ \cline{3-4}
&&$14{\rm TeV}$,~$10{\rm fb}^{-1}$& 
$T_{R}\gsim10^{8}~{\rm GeV}$
for $m_{\stau}\lsim1000{\rm GeV}$
\\ \cline{2-4}
&(ii) GUT relation
&$7{\rm TeV}$,~$1{\rm fb}^{-1}$&
$T_{R}\gsim3\times10^{7}~{\rm GeV}$
for $m_{\stau}\lsim 200{\rm GeV}$
\\ \cline{3-4}
&(Fig.~\ref{fig:ewgut})&$8{\rm TeV}$,~$5{\rm fb}^{-1}$&
$T_{R}\gsim3\times10^{7}~{\rm GeV}$
for $m_{\stau}\lsim 300{\rm GeV}$
 \\ \cline{3-4}
&&$14{\rm TeV}$,~$10{\rm fb}^{-1}$& 
$T_{R}\gsim3\times 10^{7}~{\rm GeV}$
for $m_{\stau}\lsim400{\rm GeV}$
\\ \cline{3-4}
\hline
\hline
(B) enhanced by  
&(i) $m_{\tilde{B}}=m_{\tilde{W}}=1.1m_{\stau}$ 
&$7{\rm TeV}$,~$1{\rm fb}^{-1}$&$T_{R}\gsim10^{9}~{\rm GeV}$
for $m_{\stau}\lsim700{\rm GeV}$
\\ \cline{3-4}
large ${\cal A}_{\stau\stau h}$&(Fig.~\ref{fig:11})&$8{\rm TeV}$,~$5{\rm fb}^{-1}$&
$T_{R}\gsim10^{9}~{\rm GeV}$
for $m_{\stau}\lsim1000{\rm GeV}$
\\ \cline{3-4}
&&$14{\rm TeV}$,~$10{\rm fb}^{-1}$&$T_{R}\gsim3\times 10^{8}~{\rm GeV}$
for $m_{\stau}\lsim900{\rm GeV}$ 
\\ \cline{2-4}
&(ii) GUT relation&$7{\rm TeV}$,~$1{\rm fb}^{-1}$&
$T_{R}\gsim5\times 10^{7}~{\rm GeV}$
for $m_{\stau}\lsim200{\rm GeV}$ 
\\ \cline{3-4}
&(Fig.~\ref{fig:gut})&$8{\rm TeV}$,~$5{\rm fb}^{-1}$&
$T_{R}\gsim5\times10^{7}~{\rm GeV}$
for $m_{\stau}\lsim 300{\rm GeV}$
 \\ \cline{3-4}
&&$14{\rm TeV}$,~$10{\rm fb}^{-1}$&
$T_{R}\gsim5\times 10^{7}~{\rm GeV}$
for $m_{\stau}\lsim400{\rm GeV}$ 
\\ \cline{3-4}
\hline
\hline
(C) enhanced by 
&(i) $m_{\tilde{B}}=m_{\tilde{W}}=1.1m_{\stau}$ 
&$7{\rm TeV}$,~$1{\rm fb}^{-1}$&$T_{R}\gsim5\times 10^{9}~{\rm GeV}$
\\ \cline{3-4}
large ${\cal A}_{\stau\stau H}$&(Fig.~\ref{fig:resonance11})&$8{\rm TeV}$,~$5{\rm fb}^{-1}$&
$T_{R}\gsim3\times 10^{9}~{\rm GeV}$\\ \cline{3-4}
near the pole of H&&$14{\rm TeV}$,~$10{\rm fb}^{-1}$&$T_{R}\gsim2\times 10^{9}~{\rm GeV}$ \\ \cline{2-4}
&(ii) GUT relation&$7{\rm TeV}$,~$1{\rm fb}^{-1}$&
$T_{R}\gsim2\times 10^{9}~{\rm GeV}$ for $m_{\stau}\lsim200{\rm GeV}$  
\\ \cline{3-4}
&(Fig.~\ref{fig:resonancegut})&$8{\rm TeV}$,~$5{\rm fb}^{-1}$&
$T_{R}\gsim10^{9}~{\rm GeV}$
for $m_{\stau}\lsim 250{\rm GeV}$ \\ \cline{3-4}
&&$14{\rm TeV}$,~$10{\rm fb}^{-1}$&$T_{R}\gsim2\times 10^{9}~{\rm GeV}$ \\ \cline{3-4}
\hline

\end{tabular}
\end{center}
\caption{The range of the reheating temperature with which more than 10 stau signals are expected in
each scenario.}
\label{result}
\end{table}

%%%%%%%%%%%%%%
\begin{figure}[t]
\begin{center}
\includegraphics[width=12cm]{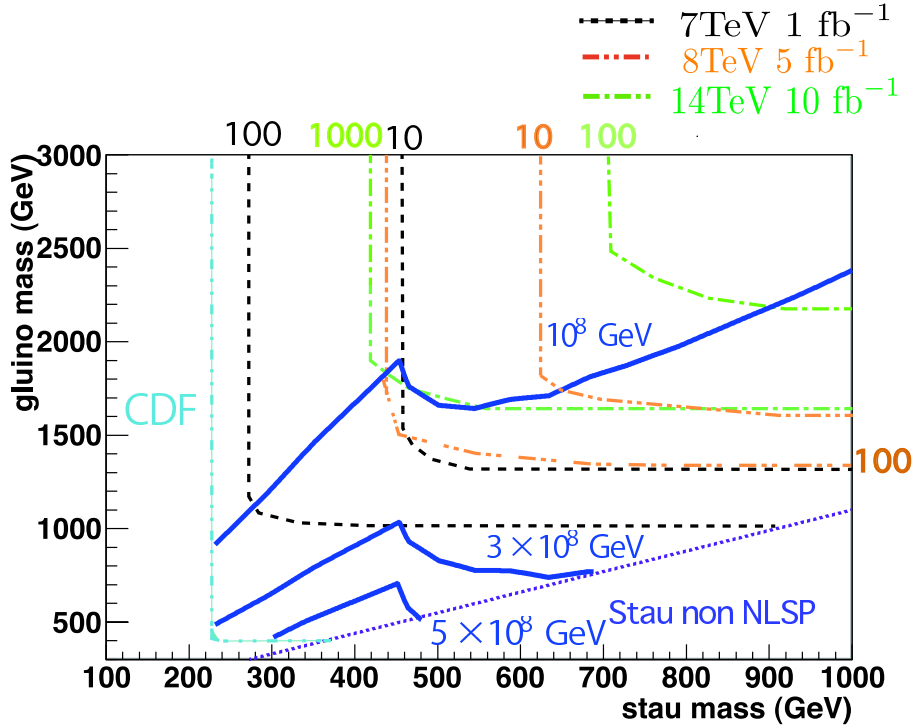}
\end{center}
\caption{Same as Fig.\ref{fig:ew11} but with $m_{ \rm squark}\simeq M_3$.}
\label{fig:squark}
\end{figure}
%%%%%%%%%%%%%%

Although we have considered
only gaugino production and stau direct production at the collisions
when discussing the Tevatron constraint and LHC signatures,
gluino-squark production becomes
dominant when the squark mass is relatively light.
We show the Tevatron constraint and LHC signatures in Fig.\ref{fig:squark} 
when the  squark mass $m_{\tilde{q}}$ equals to $m_{\tilde{g}}$
and the gaugino masses satisfy the condition (i), $m_{\tilde{B}} = m_{\tilde{W}} = 1.1 m_{\stau}$.
We also show the upper bound on the gluino mass in the case of (A)-(i).
The cross section of gluino-squark pair production is roughly ten times 
larger than the gluino pair production.
Thus, in the region where the gluino pair production is dominant in Fig.\ref{fig:ew11}, 
the expected number of stau signals  increase
in this light squark case. 
As can be seen from Fig.~\ref{fig:squark}, 
the contour lines of the constant stau number shift upwards by $\Delta m_{\tilde{g}}\simeq 
200 \sim 400~{\rm GeV}$, compared with Fig.~\ref{fig:ew11}.
On the other hand, the result does not change very much when there is the GUT relation
even for $m_{\tilde{q}}\simeq m_{\tilde{g}}$. That is because 
the gluino production is not the dominant channel of the SUSY events
in the regions around the vertical lines and the horizontal lines
 of, e.g., Fig.\ref{fig:ewgut}.
%%%%%%%%%%%%%%%%%%%%%%%%%%%%
\section{Conclusion and Discussion}
%%%%%%%%%%%%%%%%%%%%%%%%%%%%

In this paper, the observability of the stau signals at LHC has been studied 
in the scenario with the gravitino LSP and the stau NLSP under the cosmological constraints. 
It was seen that a higher reheating temperature predicts a lower gluino mass,
and therefore more stau signals can be observed at LHC. 
The upper bound on the gluino mass depends on the annihilation process of the stau,
and we have considered three cases that (A) the annihilation is dominated by the electroweak processes,
(B) the stau annihilation into the light Higgs boson is enhanced, and (C) 
the stau annihilation takes place near the pole of the heavy Higgs.
In the case (A), which is true in most of the MSSM parameter region, 
the reheating temperature cannot be larger than $7\times 10^{8}~{\rm GeV}$. 
If we further assume that the gaugino masses satisfy the GUT relation,  the upper bound becomes severer and
it cannot exceed $7\times10^7~{\rm GeV}$.
On the other hand, the reheating temperature can be much higher for the cases (B) and (C). 
Especially, we saw that the reheating temperature can be as large as 
$T_R>2\times 10^{9}~{\rm GeV}$, which is required by the thermal leptogenesis, in the case (C)
even if the gaugino masses satisfy the GUT relation.

We have then investigated the Tevatron constraints and the LHC signatures of the long-lived 
staus by imposing the trigger and the cut conditions. We found that, when the stau annihilates 
mainly via the electroweak process, it is expected that 
more than 10 stau signals are observed
at the first stage of LHC with $\sqrt{s} = 7{\rm TeV}$ and $L_i = 1{\rm fb}^{-1}$
in the region where $T_R$ is larger than $3\times 10^{8}~{\rm GeV}$.
When the heavy Higgs boson effectively contributes to the stau 
annihilation, the first stage of LHC has a sensitivity to the parameter region of $T_R>5\times 
10^{9}~{\rm GeV}$. The LHC sensitivity improves greatly as the collider energy and the luminosity 
increase. The stau signals will be detected if $T_R$ exceeds $10^{8}~{\rm GeV}$ as long as 
$m_{\stau}\lsim1000{\rm GeV}$. It is also emphasized that all the regions which satisfy $T_R>
2\times 10^{9}~{\rm GeV}$ can be checked by LHC with $\sqrt{s}=14{\rm TeV}$ and $L_i=10{\rm fb}^{-1}$.

We have also studied a specific case of the gaugino mass spectrum, i.e. the GUT relation. 
The LHC sensitivity as well as the cosmological constraint on the gluino mass is sensitive to the 
gaugino mass spectrum. The cosmological upper bound becomes severer
than the case with generic gaugino masses, while at LHC the chargino 
and/or neutralino channels dominate the productions of the SUSY events instead of the channels of 
the colored (gluino) SUSY particles, especially for a low gluino mass region. It was found that the 
LHC has a detection sensitivity for lower reheating temperature models. 

In this paper, we have assumed that there is no entropy production after the reheating epoch.
However, our results can also be applied to the case with an entropy production, 
by replacing the reheating temperature $T_R$ with $T_R^{\rm eff}=T_R/\Delta$,
where $\Delta$ is the dilution factor of the gravitino abundance.
Even if the entropy production occurs after the freeze-out of the stau and before the BBN,
the results of the case (C) (Figs.\ref{fig:resonance11} and \ref{fig:resonancegut}), where there is no BBN constraint, 
hold for $T_R^{\rm eff}$.

It is usually difficult to probe the reheating epoch of the universe directly. Nonetheless, the LHC 
has a sensitivity to the high reheating temperature models, as we discussed in this paper. Once 
heavy charged tracks will be observed at LHC, the reheating temperature receives an upper 
bound, depending on the SUSY mass spectrum. 
This enables us to reveal features of the early universe.

\section*{Acknowledgment}
The work of K.H. was supported by JSPS Grant-in-Aid for Young Scientists (B) (21740164) and
Grant-in-Aid for Scientific Research (A) (22244021).
The work of K.N. was supported by JSPS Grant-in-Aid for JSPS Fellows.
This work was supported by World Premier International Center Initiative (WPI Program),
MEXT, Japan.

\appendix
\section{Vacuum stability constraint on the stau trilinear coupling}
\label{sec:App}
%%%%%
In this appendix, the vacuum constraint on the stau trilinear coupling is discussed, which is used in Sec.~\ref{sec:B} and Sec.~\ref{sec:C}.
The lifetime of our vacuum can be estimated by using the ``bounce method" \cite{vacuum1}.
By using the euclidian action $S_E[\bar{\phi}]$ with the bounce solution $\bar{\phi}$, 
the decay rate of the false vacuum per unit volume is estimated by
\begin{eqnarray}
  \Gamma/V \simeq 
  E^4 \exp\left[ -S_E[\bar{\phi}] \right],
\end{eqnarray}
where $E$ is the typical energy scale of the potential.
We put the value of the potential energy 
at the global minimum of the potential
as 0. Since the lifetime of our vacuum must be longer than the age of the universe, 
\begin{eqnarray}
\Gamma/V\times \left(\frac{1}{H_0}\right)^4\ll 1
\end{eqnarray}
must be satisfied,
where $H_0$ is present Hubble constant. This constraint gives the lower bound on the bounce,
\begin{eqnarray}
\label{bouncelower}
S_E[\bar{\phi}]\gsim400.
\end{eqnarray}

Let us start from the case that the $\stau$-$\stau$-$h$
  trilinear coupling ${\cal A}_{\stau \stau h^0}$
is large.
In our analysis, the euclidian action takes the form of
\begin{eqnarray}
  S_E[\tilde\tau, h] = \int^{\infty}_{-\infty} d^4{\rm x}_E 
  \left[ 
    \frac{1}{2}(\partial_i\stau)(\partial_i\stau) + 
    \frac{1}{2}(\partial_i h)(\partial_i h) + 
    U(\stau, h) 
  \right],
\end{eqnarray}
where $U(\stau,h)$ is the potential of $\stau$ and $h$. We calculate the potential at one loop level for Higgs potential and at tree level for $\stau$ potential.  Now the constraint (\ref{bouncelower}) gives the upper bound on ${\cal A}_{\stau \stau h^0}$ for a given stau mass which is shown in Fig.\ref{fig:ccbbound}  of Sec.~\ref{sec:B}, for $m_h=120$~GeV.
When we calculate the bounce, we take straight line path from the false vacuum to the true vacuum in $(\stau,h)$ plane. We checked that the full analysis using two dimensional path 
changes the result from the straight line approximation by only ${\cal O}(1)\%$.

We also have to care about the thermal transition of our vacuum to CCB vacuum in the early stage of the universe \cite{ccbbound}. The procedure of calculating the thermal transition rate is similar to that of zero temperature case. We use free energy $F$ instead of euclidian action $S_E$
to calculate the bounce solution $\bar{\phi}$. 
The thermal transition rate at temperature $T$ per unit volume is given by
\begin{eqnarray}
\Gamma(T)/V\simeq T^4 e^{-F[\bar{\phi},T]/T}.
\end{eqnarray}
Since the thermal transition must not occur,
\begin{eqnarray}
\label{thermaltransition}
\int^{t_f}_{t_i}dt\frac{1}{H(t)^3}(\Gamma(T)/V)\ll 1
\end{eqnarray}
must be satisfied,
where $t_i$ and $t_f$ are initial and final time respectively, and $H(t)$ is the
Hubble constant at each time.
In our
analysis, the free energy takes the form of
\begin{eqnarray}
  F[\tau, h,T] = 
  \int^{\infty}_{-\infty} d^3 x 
  \left[
    \frac{1}{2}(\partial_i\stau)(\partial_i\stau) + 
    \frac{1}{2}(\partial_i h)(\partial_i h) + 
    U(\stau,h)+
    \delta V_{th}(\stau, h, T)
  \right],
\end{eqnarray}
where $\delta V_{th}$ is the thermal potential calculated by thermal field theory. 
The thermal potential include the contribution from top quark and 
gauge bosons of U(1) and SU(2) gauge symmetries in the Standard Model. 
From the constraint (\ref{thermaltransition}), 
We get the upper bound on ${\cal A}_{\stau \stau h^0}$. 
The resultant upper bound on ${\cal A}_{\stau \stau h^0}$ for a given stau mass is also shown
in Fig.\ref{fig:ccbbound}.

The upper bound on the $\stau$-$\stau$-$H$  trilinear coupling
${\cal A}_{\stau\stau H}$, which is discussed in Sec.~\ref{sec:C}, 
can be obtained in the similar way.
The result is shown in Fig.~\ref{fig:CCBlarge} for $m_H = 2m_{\stau}$,
where we take account of the Higgs potential at tree level, and
the contributions from bottom quark, tau lepton and the gauge bosons of the U(1) and
SU(2) gauge symmetries for the thermal potential.


\begin{thebibliography}{99}

\bibitem{LHCCHAMP}
%\cite{Khachatryan:2011ts}
%\bibitem{Khachatryan:2011ts}
  V.~Khachatryan {\it et al.} [ CMS Collaboration ],
  %``Search for Heavy Stable Charged Particles in pp collisions at sqrt(s)=7 TeV,''
  JHEP {\bf 1103 } (2011)  024.
  [arXiv:1101.1645 [hep-ex]];
%\cite{Aad:2011yf}
%\bibitem{Aad:2011yf}
  G.~Aad {\it et al.} [ ATLAS Collaboration ],
  %``Search for stable hadronising squarks and gluinos with the ATLAS experiment at the LHC,''  
  [arXiv:1103.1984 [hep-ex]].
  



\bibitem{FIY}
  M.~Fujii, M.~Ibe and T.~Yanagida,
  %``Upper bound on gluino mass from thermal leptogenesis,''
  Phys.\ Lett.\  B {\bf 579}, 6 (2004)
  [arXiv:hep-ph/0310142].
  %%CITATION = PHLTA,B579,6;%%
%\cite
%\cite{Endo:2010ya}

\bibitem{RandS}
%\cite{Roszkowski:2004jd}
%\bibitem{Roszkowski:2004jd}
  L.~Roszkowski, R.~Ruiz de Austri and K.~Y.~Choi,
  %``Gravitino dark matter in the CMSSM and implications for leptogenesis  and
  %the LHC,''
  JHEP {\bf 0508} (2005) 080
  [arXiv:hep-ph/0408227];
  %%CITATION = JHEPA,0508,080;%%%\cite{Pradler:2006qh}
%\bibitem{Pradler:2006qh}
 J.~Pradler and F.~D.~Steffen,
 %``Thermal Gravitino Production and Collider Tests of Leptogenesis,''
 Phys.\ Rev.\  D {\bf 75} (2007) 023509
 [arXiv:hep-ph/0608344];
 %%CITATION = PHRVA,D75,023509;%%
%\cite{Choi:2007rh}
%\bibitem{Choi:2007rh}
  K.~Y.~Choi, L.~Roszkowski and R.~Ruiz de Austri,
  %``Determining Reheating Temperature at Colliders with Axino or Gravitino Dark
  %Matter,''
  JHEP {\bf 0804} (2008) 016
  [arXiv:0710.3349 [hep-ph]];
  %%CITATION = JHEPA,0804,016;%%
%\cite{Steffen:2008bt}
%\bibitem{Steffen:2008bt}
  F.~D.~Steffen,
  %``Probing the Reheating Temperature at Colliders and with Primordial
  %Nucleosynthesis,''
  Phys.\ Lett.\  B {\bf 669} (2008) 74
  [arXiv:0806.3266 [hep-ph]].
  %%CITATION = PHLTA,B669,74;%%


%\cite{Fukugita:1986hr}
\bibitem{Fukugita:1986hr}
  M.~Fukugita, T.~Yanagida,
  %``Baryogenesis Without Grand Unification,''
  Phys.\ Lett.\  {\bf B174 } (1986)  45.

\bibitem{lepto}  
%\cite{Buchmuller:2005eh}
%\bibitem{Buchmuller:2005eh}
  W.~Buchmuller, R.~D.~Peccei, T.~Yanagida,
  %``Leptogenesis as the origin of matter,''
  Ann.\ Rev.\ Nucl.\ Part.\ Sci.\  {\bf 55}, 311-355 (2005).
  [hep-ph/0502169];
%\cite{Davidson:2008bu}
%\bibitem{Davidson:2008bu}
  S.~Davidson, E.~Nardi, Y.~Nir,
  %``Leptogenesis,''
  Phys.\ Rept.\  {\bf 466}, 105-177 (2008).
  [arXiv:0802.2962 [hep-ph]].




\bibitem{Endo:2010ya}
  M.~Endo, K.~Hamaguchi and K.~Nakaji,
  %``Probing High Reheating Temperature Scenarios at the LHC with Long-Lived
  %Staus,''
  JHEP {\bf 1011}, 004 (2010)
  [arXiv:1008.2307 [hep-ph]].
  %%CITATION = JHEPA,1011,004;%%


\bibitem{enhance1}
 M.~Ratz, K.~Schmidt-Hoberg and M.~W.~Winkler,
 %``A note on the primordial abundance of stau NLSPs,''
 JCAP {\bf 0810}, 026 (2008)
 [arXiv:0808.0829 [hep-ph]].
%%CITATION = JCAPA,0810,026;\bibitem{enhance1}

\bibitem{enhance2}
 J.~Pradler and F.~D.~Steffen,
 %``Thermal relic abundances of long-lived staus,''
 Nucl.\ Phys.\  B {\bf 809}, 318 (2009)
 [arXiv:0808.2462 [hep-ph]].
 %%CITATION = NUPHA,B809,318;%%%\cite{Ratz:2008qh}%%



\bibitem{thermaltr}
%\cite{Moroi:1993mb}
%\bibitem{Moroi:1993mb}
  T.~Moroi, H.~Murayama and M.~Yamaguchi,
  %``Cosmological constraints on the light stable gravitino,''
  Phys.\ Lett.\  B {\bf 303}, 289 (1993).
  %%CITATION = PHLTA,B303,289;%%
%\cite{Bolz:2000fu}
\bibitem{Bolz:2000fu}
  M.~Bolz, A.~Brandenburg and W.~Buchmuller,
  %``Thermal Production of Gravitinos,''
  Nucl.\ Phys.\  B {\bf 606}, 518 (2001)
  [Erratum-ibid.\  B {\bf 790}, 336 (2008)]
  [arXiv:hep-ph/0012052].
  %%CITATION = NUPHA,B606,518;%%
%\cite{Pradler:2006hh}
\bibitem{Pradler:2006hh}
  J.~Pradler and F.~D.~Steffen,
  %``Constraints on the reheating temperature in gravitino dark matter
  %scenarios,''
  Phys.\ Lett.\  B {\bf 648}, 224 (2007)
  [arXiv:hep-ph/0612291].
  %%CITATION = PHLTA,B648,224;%%
%\cite{Rychkov:2007uq}
\bibitem{Rychkov:2007uq}
  V.~S.~Rychkov, A.~Strumia,
  %``Thermal production of gravitinos,''
  Phys.\ Rev.\  {\bf D75}, 075011 (2007).
  [hep-ph/0701104].
  
\bibitem{inflatondecay}
  M.~Endo, F.~Takahashi and T.~T.~Yanagida,
  %``Inflaton Decay in Supergravity,''
  Phys.\ Rev.\  D {\bf 76} (2007) 083509
  [arXiv:0706.0986 [hep-ph]], and references therein.
  %%CITATION = PHRVA,D76,083509;%%
  
\bibitem{modulidecay}
  M.~Endo, K.~Hamaguchi and F.~Takahashi,
  %``Moduli-induced gravitino problem,''
  Phys.\ Rev.\ Lett.\  {\bf 96}, 211301 (2006)
  [arXiv:hep-ph/0602061];
  %\cite{Nakamura:2006uc}
%\bibitem{Nakamura:2006uc}
  S.~Nakamura, M.~Yamaguchi,
  %``Gravitino production from heavy moduli decay and cosmological moduli problem revived,''
  Phys.\ Lett.\  {\bf B638 } (2006)  389-395.
  [hep-ph/0602081].

  
%\cite{Amsler:2008zzb}
\bibitem{darkmatterdensity}
 C.~Amsler {\it et al.}  [Particle Data Group],
 %``Review of particle physics,''
 Phys.\ Lett.\  B {\bf 667} (2008) 1.
 %%CITATION = PHLTA,B667,1;%%


\bibitem{cbbn1}
 M.~Pospelov,
 %``Particle physics catalysis of thermal big bang nucleosynthesis,''
 Phys.\ Rev.\ Lett.\  {\bf 98}, 231301 (2007)
 [arXiv:hep-ph/0605215];
 %%CITATION = PRLTA,98,231301;%%
%\cite{Pradler:2008qc}
%\cite{Kamimura:2008fx}
%\bibitem{Kamimura:2008fx}
  M.~Kamimura, Y.~Kino, E.~Hiyama,
  %``Big-Bang Nucleosynthesis Reactions Catalyzed by a Long-Lived Negatively-Charged Leptonic Particle,''
  Prog.\ Theor.\ Phys.\  {\bf 121 } (2009)  1059-1098.
  [arXiv:0809.4772 [nucl-th]], and references therein.

%\cite{Kawasaki:2008qe}
\bibitem{bbn2}
 M.~Kawasaki, K.~Kohri, T.~Moroi and A.~Yotsuyanagi,
 %``Big-Bang Nucleosynthesis and Gravitino,''
 Phys.\ Rev.\  D {\bf 78}, 065011 (2008)
 [arXiv:0804.3745 [hep-ph]], and references therein.
 %%CITATION = PHRVA,D78,065011;%%
%\cite{Asaka:2000zh}

\bibitem{Asaka:2000zh}
  T.~Asaka, K.~Hamaguchi and K.~Suzuki,
  %``Cosmological gravitino problem in gauge mediated supersymmetry breaking
  %models,''
  Phys.\ Lett.\  B {\bf 490}, 136 (2000)
  [arXiv:hep-ph/0005136].
  %%CITATION = PHLTA,B490,136;%%
  
  %\cite{Kusenko:1996jn}
     \bibitem{ccbbound}
       A.~Kusenko, P.~Langacker and G.~Segre,
       %``Phase Transitions and Vacuum Tunneling Into Charge and Color Breaking
       %Minima in the MSSM,''
       Phys.\ Rev.\  D {\bf 54}, 5824 (1996)
       [arXiv:hep-ph/9602414].
       %%CITATION = PHRVA,D54,5824;%%
%\cite{Sjostrand:2006za}
\bibitem{vacuum1}
%\cite{Coleman:1977py}
%\bibitem{Coleman:1977py}
  S.~R.~Coleman,
  %``The Fate Of The False Vacuum. 1. Semiclassical Theory,''
  Phys.\ Rev.\  D {\bf 15}, 2929 (1977)
  [Erratum-ibid.\  D {\bf 16}, 1248 (1977)];
  \\
  %%CITATION = PHRVA,D15,2929;%%
%\cite{Callan:1977pt}
%\bibitem{Callan:1977pt}
  C.~G.~Callan and S.~R.~Coleman,
  %``The Fate Of The False Vacuum. 2. First Quantum Corrections,''
  Phys.\ Rev.\  D {\bf 16}, 1762 (1977).
  %%CITATION = PHRVA,D16,1762;%%
%\cite{Griest:1990kh}
\bibitem{CMBdistortion}
  L.~Roszkowski, R.~Ruiz de Austri and K.~Y.~Choi,
  %``Gravitino dark matter in the CMSSM and implications for leptogenesis  and
  %the LHC,''
  JHEP {\bf 0508}, 080 (2005)
  [arXiv:hep-ph/0408227]
  %%CITATION = JHEPA,0508,080;%%
%\cite{}
%\cite{Belanger:2010gh}
\bibitem{micromega}
  G.~Belanger, F.~Boudjema, P.~Brun, A.~Pukhov, S.~Rosier-Lees, P.~Salati, A.~Semenov,
  %``Indirect search for dark matter with micrOMEGAs2.4,''
  Comput.\ Phys.\ Commun.\  {\bf 182}, 842-856 (2011).
  [arXiv:1004.1092 [hep-ph]].


\bibitem{CMBdistort}
See, for instance,  L.~Roszkowski, R.~Ruiz de Austri and K.~Y.~Choi, in~\cite{RandS} and references therein.


\bibitem{pythia}
 T.~Sjostrand, S.~Mrenna and P.~Z.~Skands,
 %``PYTHIA 6.4 Physics and Manual,''
 JHEP {\bf 0605}, 026 (2006)
 [arXiv:hep-ph/0603175].
 %%CITATION = JHEPA,0605,026;%%
%\cite{Beenakker:1996ed}
\bibitem{prospino}
  W.~Beenakker, R.~Hopker and M.~Spira,
  %``PROSPINO: A program for the PROduction of Supersymmetric Particles In
  %Next-to-leading Order QCD,''
  arXiv:hep-ph/9611232.
  %%CITATION = HEP-PH/9611232;%%
%\cite{:2003hw}

\bibitem{pgs}
The information on Pretty Good Simulation of high energy collisions (PGS4) can be seen in
http://www.physics.ucdavis.edu/\verb|%|7Econway/research/research.html.

\bibitem{tevatron}
  T.~Aaltonen {\it et al.}  [CDF Collaboration],
  %``Search for Long-Lived Massive Charged Particles in 1.96 TeV $\bar{p}p$
  %Collisions,''
  Phys.\ Rev.\ Lett.\  {\bf 103}, 021802 (2009)
  [arXiv:0902.1266 [hep-ex]].
  %%CITATION = PRLTA,103,021802;%%
  
%\cite{Abazov:2008qu}
\bibitem{Abazov:2008qu}
  V.~M.~Abazov {\it et al.}  [D0 Collaboration],
  %``Search for Long-Lived Charged Massive Particles with the D0 Detector,''
  Phys.\ Rev.\ Lett.\  {\bf 102}, 161802 (2009)
  [arXiv:0809.4472 [hep-ex]].
  %%CITATION = PRLTA,102,161802;%%

%\cite{Aad:2009wy}
\bibitem{Aad:2009wy}
  G.~Aad {\it et al.}  [The ATLAS Collaboration],
  %``Expected Performance of the ATLAS Experiment - Detector, Trigger and
  %Physics,''
  arXiv:0901.0512 [hep-ex].
  %%CITATION = ARXIV:0901.0512;%%
 \bibitem{reconstruction}
S.~Tarem, S.~Bressler, H.~Nomoto and A.~Di Mattia,
%``Trigger and reconstruction for heavy long-lived charged particles with the
%ATLAS detector,''
Eur.\ Phys.\ J.\  C {\bf 62}, 281 (2009).
%%CITATION = EPHJA,C62,281;%%
%cite{Djouadi:2006bz}
%cite{Aaltonen:2009kea}


\end{thebibliography}
\end{document}